\documentclass[journal,9pt]{IEEEtran}
\usepackage{graphicx,subfigure}
\usepackage{amsfonts,amsmath,amssymb, mathrsfs, amsthm, cite}
\usepackage{multirow}
\usepackage{epic,eepic,eepicemu}
\usepackage{epsf}
\usepackage{epsfig}
\usepackage{graphics,pgfplots}
\usepackage{array} 
\usepackage{mathrsfs}
\usepackage{psfrag}
\usepackage{tikz}
\usepackage{url}
\usepackage{enumerate}
\usepackage{booktabs}
\usetikzlibrary{arrows}
\usepackage{verbatim}

\usepackage{bm}
\usepackage{tikz,amsmath,graphicx,xcolor,soul,verbatim}
\definecolor{mycolor1}{rgb}{0.00000,0.44700,0.74100}
\ifodd 1 
\else  \fi

\usetikzlibrary{arrows}
\usepackage{verbatim}

\newtheorem{definition}{Definition}
\newtheorem{example}{Example}
\newtheorem{lemma}{Lemma}

\newtheorem{proposition}{Proposition}
\newcommand*{\rom}[1]{\expandafter\@slowromancap\romannumeral #1@}
\makeatother

\usepackage[skip=2pt,font=footnotesize]{caption}
\newcommand{\abs}[1]{\lvert#1\rvert}

\newcommand{\be}{\begin{equation}}
\newcommand{\ee}{\end{equation}}
\newcommand{\ben}{\begin{equation*}}
\newcommand{\een}{\end{equation*}}
\newcommand{\mc}{\mathcal}

\newcommand{\bsym}{\boldsymbol}

\newcommand{\Bc}{\mathcal{B}}
\newcommand{\Cc}{\mathcal{C}}

\newcommand{\Gc}{\mathcal{G}}
\newcommand{\Lc}{\mathcal{L}}
\newcommand{\Tc}{\mathcal{T}}
\newcommand{\Vc}{\mathcal{V}}

\newcommand{\syn}{\textsf{syn}}
\newcommand{\Ab}{\bm{A}}

\DeclareMathSymbol{:}{\mathbin}{operators}{"3A}

\IEEEoverridecommandlockouts

\begin{document}

%%%%%%%%%%%%%%%%%%%%%%%%%%%%%%%

%%%%%%%%%%%%%%%%%%%%%%%%%%%%%%%
\title{%\hspace{-2.2mm}
Multilayer Codes for Synchronization\\ from Deletions and Insertions}
%%%%%%%%%%%%%%%%%%%%%%%%%%%%%%%

%%%%%%%%%%%%%%%%%%%%%%%%%%%%%%%

\author{Mahed Abroshan, Ramji Venkataramanan and Albert Guill\'en i F\`abregas\thanks{M. Abroshan is with the Alan Turing Institute (mabroshan@turing.ac.uk). This work was completed when he was at the Department of Engineering, University of Cambridge.}
\thanks{ R. Venkataramanan is with the Department of Engineering, University of Cambridge, UK (ramji.v@eng.cam.ac.uk).}
\thanks{A. Guill\'en i F\`abregas is with the Department of Information and Communication
Technologies, Universitat Pompeu Fabra, Barcelona 08018, Spain,
also with the Instituci\'o Catalana de Recerca i Estudis Avan\c{c}ats (ICREA),
Barcelona 08010, Spain, and also with the Department of Engineering, University
of Cambridge, UK (e-mail: guillen@ieee.org).}
\thanks{This paper was presented in part in the 2017 IEEE Information Theory Workshop, Kaohsiung, Taiwan, Nov. 2017.}
\thanks{This work was supported in part by the European Research Council under Grants 725411, in part by the Spanish Ministry of Economy and Competitiveness under Grant  TEC2016-78434-C3-1-R.}
}
%\date{\today}

\maketitle

\begin{abstract}
Consider two remote nodes (encoder and decoder), each with a binary sequence. The encoder's sequence $X$  differs from the decoder's sequence $Y$  by a small number of edits (deletions and insertions). The goal is to construct a message $M$, to be sent via a one-way error free link, such that the decoder can reconstruct $X$ using $M$ and $Y$. In this paper, we devise a coding scheme for this  one-way synchronization model. The scheme is based on multiple layers of Varshamov-Tenengolts (VT) codes combined with off-the-shelf linear error-correcting codes, and uses a list decoder.  We bound the expected list size of the decoder under certain assumptions, and validate its performance via numerical simulations. We also consider an alternative decoder that uses only the constraints from the VT codes (i.e., does not require a linear code), and  has a smaller redundancy at the expense of a slightly larger average list size.  
\end{abstract}

\section{Introduction}
Consider two remote nodes with binary sequences $X$ and $Y$, respectively. The sequence $Y$ is an \emph{edited} version of $X$, where the edits consist of deletions and insertions.   In the synchronization model shown in Fig. \ref{Source-model}, the node with $X$ (the encoder) sends a message $M$ via an error-free link to the other node (the decoder), which attempts to reconstruct $X$ using $M$ and $Y$. The goal is to design a scheme  so that the decoder can reconstruct $X$ with minimal communication, i.e., we want to minimize the number of bits used to represent the message $M$. 

This synchronization model is relevant in a number of  applications including distributed file editing, and systems for file backup and sharing (e.g., Dropbox). The synchronization problem has been studied in several previous works, both in the one-way setting \cite{Orlitsky93,irmak2005improved,MaKTse11,belazzougui2016edit,cheng2018,haeupler2018optimal,jowhari,hanna2019guess}, and in the two-way setting where the encoder and decoder can exchange multiple rounds of messages \cite{CormodePSV00,OrlitskyV01,RVSync15,YazdiDol14,sala2016synchronizing}. Some practical file synchronization tools such as  rsync \cite{trigdell99} also use multiple rounds of information exchange. We discuss the prior work on one-way synchronization  in more detail in Section \ref{subsec:rel_work}.

We seek codes for {one-way} synchronization: in Fig. \ref{Source-model},  the message $M$ is produced by the encoder using  only $X$, with no knowledge of $Y$, except that the number of edits is at most $k$. We assume that the decoder knows the length of $X$, which is denoted by $n$.   The message $M$ belongs to a finite set $\mc{M}$ with cardinality $\abs{\mc{M}}$. The synchronization rate (or redundancy per symbol) is defined as $R = \frac{ \log \abs{\mc{M}}}{n}$. (Throughout the paper, $\log$ denotes logarithm with base 2.) We would like to design a code for reliable synchronization with $R$ as small as possible, noting that $R=1$ is equivalent to the encoder sending the entire string $X$.

In this paper, we construct a code based on multiple layers of Varshomov-Tenengolts (VT) codes \cite{VT65}, 
for synchronization when the number of edits  $k$ is small compared to $n$. The output of the decoder is a small list of sequences that is guaranteed to contain the correct sequence $X$. We observe from simulations that with a careful choice of the code parameters, the list size rarely exceeds $2$ or $3$; for reasonably large $n$, the list size can be made $1$, i.e., $X$ is exactly reconstructed. For example, we construct a code of length $n=378$ that can synchronize from $k=7$ edits with $R=0.365$, and a length $n=2800$ code which can synchronize from $k=10$ edits with $R=0.135$. (Details in Section \ref{sec:simulation} and \ref{sec:ins:dec}.)

%which are known to be efficient single-edit correcting codes, 

\begin{figure}[t]
\begin{center}
\begin{tikzpicture}[node distance=3.5cm,auto,>=latex', scale=1.4]
    \draw[black, thick,rounded corners] (0,0) rectangle (1,0.8) node[font=\fontsize{9}{144}\selectfont,pos=.5,anchor=center] {Encoder};
    \draw[black, thick,->] (1,.3) -- (3,.3) node[font=\fontsize{9}{144}\selectfont,pos=.5,anchor=south] {$M=\mathcal E(X)$}; 
    \draw[black,thick,rounded corners](3,0) rectangle (4,0.8) node[font=\fontsize{9}{144}\selectfont,pos=.5,anchor=center]{Decoder}; 
    \draw[black, thick,->] (-.7,0.4) ->(0,0.4) node[font=\fontsize{8}{144}\selectfont,pos=.2,anchor=south]{$X$};
    \draw[black, thick,->] (3.5,1.3) -> (3.5,.8) node[font=\fontsize{8}{144}\selectfont,pos=.2,above=1pt] {$Y$};
    \draw[black, thick,->] (4,0.4) -> (4.7,.4) node[font=\fontsize{8}{144}\selectfont,pos=.85,above=1 pt] {$\hat{X}$};
\end{tikzpicture}
\end{center}
\caption{Synchronization Model}\label{Source-model}
\vspace{-8pt}
\end{figure}

To  explain the main ideas in the code construction and the decoding algorithm, we largely focus on the case where the edits are all deletions.  Section \ref{sec:ins:dec} describes how to modify the decoder (keeping the same encoding  scheme) to reconstruct a combination of insertions and deletions. 

\subsection{Overview of the code construction}

The starting point for our code construction is the family of Varshamov-Tenengolts (VT) codes \cite{VT65,Lev65}. Each VT code is a single edit correcting code, where the edit can be either an insertion or a deletion. A construction based on modifications of VT codes is also used for recovering a burst of consecutive deletions or insertions \cite{schoeny2017codes}. As observed in \cite{Orlitsky93},  the single edit correcting VT code provides an elegant scheme for synchronizing from a single edit: the encoder simply sends the VT syndrome  of the sequence $X$. The VT syndrome (defined in the next section) indicates which VT code $X$ belongs to. The decoder then uses the single edit correcting property  of the VT code to recover $X$.  

In our case, the code needs to synchronize from $k>1$ edits, assumed for now to be all deletions.  The encoder sends the VT syndromes of various substrings of $X$ to the decoder. Specifically, the length $n$ sequence $X$ is divided into smaller chunks of $n_c$ bits each. The encoder then computes VT syndromes for two kinds of substrings: \emph{blocks} which are composed of adjacent chunks, and \emph{chunk-strings} which are composed of well-separated chunks.  Fig. \ref{fig_enc} shows an example where  $X$ of length $12$ is divided into $4$ length-$3$ chunks. The blocks $B_1$ and $B_2$ are  each formed by combining two adjacent chunks, while the chunk-strings $C_1$ and $C_2$ are each formed by combining two alternate chunks. In this case, the encoder sends the VT syndromes of $B_1, B_2, C_1$, and $C_2$.  

The intersecting VT constraints of blocks and chunk-strings help the decoder to estimate locations of  the edits. The VT syndromes serve a dual purpose: i) they are used to recover deleted bits in blocks or chunk-strings inferred to have a single deletion, and this recovery may result in new blocks and chunk-strings with a single  deletion; ii)  the VT syndromes also act as hashes that eliminate a large number of potential deletion patterns,  allowing the decoder to localize the deletions to a relatively small set of chunks. 

The  final part of the message  is the syndrome of $X$ with respect to the parity-check constraints of a linear code. The linear parity-check constraints are used to recover the deletions in chunks that still remain unresolved at the decoder after processing the intersecting VT constraints. We call this code construction   a two-layer code as the chunks are combined to form two kinds of intersecting substrings. The construction can be generalized to combine chunks in multiple ways to form many layers of intersecting substrings. (A three-layer construction is briefly discussed in Section \ref{sec:conc}.) Increasing the number of constraints in the code  improves its synchronization capability at the cost of increasing the redundancy.

\begin{figure}
\begin{center}
\begin{tikzpicture}[node distance=3.5cm,auto,>=latex', scale=1.4]
    \draw[black, thick] (0,0) rectangle (1.5,0.6) node[font=\fontsize{9}{144}\selectfont,pos=.5,anchor=center] {$C^1_1=x_1x_2x_3$};
    \draw[black, thick] (1.5,0) rectangle (3,0.6) node[font=\fontsize{9}{144}\selectfont,pos=.5,anchor=center] {$C^1_2=x_4x_5x_6$};
    \draw[black, thick] (3,0) rectangle (4.5,0.6) node[font=\fontsize{9}{144}\selectfont,pos=.5,anchor=center] {$C^2_1=x_7x_8x_9$};
    \draw[black,thick](4.5,0) rectangle (6.25,0.6) node[font=\fontsize{9}{144}\selectfont,pos=.5,anchor=center]{$C^2_2=x_{10}x_{11}x_{12}$};

    \draw[black, thick] (.75,0) -- (1,-.5) node[font=\fontsize{9}{144}\selectfont,pos=1,anchor=north] {$C_1=x_1x_2x_3x_7x_8x_9$};
\draw[black, thick] (3.75,0) -- (1,-0.5);    
     
\draw[black, thick] (2.25,0) -- (5,-.5) node[font=\fontsize{9}{144}\selectfont,pos=1,anchor=north] {$C_2=x_4x_5x_6x_{10}x_{11}x_{12}$};
\draw[black, thick] (5.25,0) -- (5,-0.5);    
    \draw[black, thick] (.75,0.6) -- (1.5,1) node[font=\fontsize{8}{144}\selectfont,pos=1,anchor=south] {$B_1=x_1x_2x_3x_4x_5x_6$};
    \draw[black, thick] (2.25,0.6) -- (1.5,1);
    \draw[black, thick] (3.75,0.6) -- (4.5,1) node[font=\fontsize{8}{144}\selectfont,pos=1,anchor=south] {$B_2=x_7x_8x_9x_{10}x_{11}x_{12}$};
    \draw[black, thick] (5.25,0.6) -- (4.5,1);
\end{tikzpicture}
\end{center}
\caption{Blocks and chunk-strings structure for the example where $l_1=l_2=2$}
\label{fig_enc}
\vspace{-8pt}
\end{figure} 

For decoding, we use a list decoder. The output of the decoder is the list of all length $n$ sequences that can be obtained by inserting $k$ bits into sequence $Y$, and satisfy the VT constraints and the parity-check constraints that are imposed via message $M$. The correct sequence $X$ is always in the list.

\subsection{Related work} \label{subsec:rel_work}

In \cite{irmak2005improved}, Irmak et al. propose a one-way  randomized scheme that synchronizes with a message of length $O(k \log^2 n)$, where $k$ is the number of edits and $n$ is the length of $X$. The scheme in \cite{irmak2005improved} uses a multi-level message formed by splitting $X$ into successively smaller blocks. The message at each level is computed by applying a hash function to the blocks at that level.  In a series of recent papers \cite{belazzougui2016edit,cheng2018,haeupler2018optimal,jowhari},  variants of the  construction in \cite{irmak2005improved} have been used  to  achieve synchronization with message lengths of smaller order. The deterministic scheme proposed in  \cite{cheng2018} uses a message of length $O(k \log^2 \frac{n}{k})$, and the randomized scheme  in \cite{haeupler2018optimal}  achieves synchronization with high probability with a message of length $O(k \log \frac{n}{k})$, which is the optimal order \cite{Orlitsky93}. 

The goal in these works is to obtain a synchronization scheme that is \emph{order-optimal}, i.e., a scheme  with message length (redundancy) of order close to $k \log \frac{n}{k}$ and polynomial-time encoding and decoding. The constants in these  results are not explicitly specified and can be quite large. For example, the message length for the scheme in \cite{cheng2018}  is at least $200 k\log n$ \cite{sima2019optimal}, which implies that we need $n$ to be at least a few tens of thousands before the per-symbol redundancy is less than 1. In contrast, we are interested in practical codes to synchronize from a few edits in sequences that are a few hundred to a few thousand bits long. 

From this perspective, the most relevant work to our setup is the 
``guess-and-check"  (GC) code recently proposed by Hanna and El Rouayheb in \cite{hanna2019guess}.  In the GC code, the length $n$ sequence $X$ is divided into chunks of $\log n$ bits each. Assume that $n$ is a power of 2, so that each chunk can be considered as a symbol over the field $GF(n)$. The encoder's message consists of  $c$ parity-check symbols of a systematic MDS code over $GF(n)$, computed  with the information sequence $X$.  Here $c >2k$, where $k$ is the  number of deletions. The decoder considers each pattern of $k$ deletions,  and checks whether the pattern is consistent with the parity-check symbols. Decoding is successful if there is a unique sequence consistent with all the  linear parity-check constraints. It is shown in \cite{hanna2019guess} that the probability of successful decoding is $O(n^{-(c-2k)}/ (\log n)^k )$. A list decoder for the GC code was recently considered in \cite{GClist19}.

Our construction can be viewed as a generalization of GC code. Like the GC code, we divide the sequences into chunks and use parity-check symbols as part of the message. However, the set of syndromes of intersecting VT constraints  is an essential ingredient in our construction  that is not present in the GC code. The VT constraints significantly reduce the decoding complexity  by localizing and correcting a large number of deletions, and reduces the number of parity-check symbols required.  The parity-check symbols help to recover a small number of chunks in the original string, with the large majority of chunks being resolved using the intersecting VT constraints. In fact, in Section  \ref{sec:alt-dec} we discuss a variant of the code that does not use any parity-check constraints.   The decoding complexity of our scheme is compared in detail with the GC code in Section \ref{subsec:GC_comp}.

List decoding of codes for insertions and deletions was recently analyzed in \cite{wachter2018list}. Specifically, that paper obtains a lower bound  for the maximum list size when the code consists of a single VT constraint, and shows that the list size can grow exponentially with the number of deletions. In contrast, our construction uses multiple intersecting VT constraints, and is therefore  challenging to analyze rigorously. 
 
The problem of one-way synchronization from $k$ deletions is closely related to the problem of communicating over a deletion channel that deletes $k$ bits from a length $n$ codeword \cite{OrlitskyV03}.  Constructing efficient codes for the deletion channel is known to be a challenging problem, see e.g., \cite{DaveyMackay01, abdelPFC12,brakensiekVZ16}. Any one way synchronization scheme directly yields a deletion correcting code. Indeed, for a fixed message $M$ from the synchronization scheme, one can take the codebook to be the set of all sequences for which the synchronization scheme produces $M$. Using this method, we obtain a
deletion correcting code corresponding to each message of the synchronization scheme.
However, it may not be possible to translate a deletion code directly to a synchronization scheme. For example, an efficient $k$-deletion correcting channel code  with near-optimal redundancy was recently proposed in \cite{sima2019optimal}. This code has  redundancy of $8k \log n+o(\log n)$ and its decoding complexity is $O(n^{2k+1})$. However, this code cannot be  directly translated to the one-way synchronization model since the VT-like syndrome used in the code only works for sequences with no consecutive ones. Similarly, other practical codes for deletion channel such as watermark codes \cite{DaveyMackay01} use codewords with a special structure designed to aid decoding. These codes cannot be directly applied to the one-way synchronization model where the sequence available at the encoder is arbitrary and will not have the desired structure in general. Our construction is based on VT codes because they can be  translated to a synchronization scheme for one-deletion via the VT syndromes \cite{Orlitsky93}. Designing synchronization schemes based on multiple deletion correcting channel codes is an interesting direction for future work.

\subsection{Contributions}
The organization and the contributions of the paper are as follows.
\begin{itemize}

\item  The construction of the two layer code and the encoding are described in Section \ref{sec:encoding}, and the list decoding algorithm in Section \ref{sec:decoding}. The performance of the list decoder is evaluated using numerical simulations in Section \ref{sec:simulation}.

\item In Section \ref{sec:listsize}, we obtain a  bound on the expected list size under certain assumptions.  Though not  tight, the bound gives insight into how the various code parameters affect the list size. 

\item In Section \ref{sec:dec_complexity}, we analyze the complexity of encoding and decoding. The list decoder consists of multiple steps, and the complexity of each step depends on the list size at the end of the previous step. For a fixed number of edits, the encoding complexity is linear and the decoding complexity is $O(n^3)$. However, this is based on a worst-case analysis that does not consider the effect of the VT constraints in reducing the list size. Our numerical experiments indicate  that the decoding complexity is typically  much lower. In Section \ref{subsec:GC_comp}, we compare the decoding complexity with that of the Guess and Check code via a numerical example.

\item In Section \ref{sec:alt-dec}, we discuss an alternative decoder that does not require the parity-check constraints. Eliminating these constraints reduces per-symbol redundancy at the expense of a slightly larger average list size. 

\item In Section \ref{sec:ins:dec}, we extend the decoding algorithm to handle a combination of deletions and insertions. Section \ref{sec:conc} concludes the paper with a discussion of how the two-layer construction can be generalized to multiple layers.
\end{itemize}

Before we proceed, we emphasize that the code construction and its analysis throughout the paper is for the case where the number of edits $k$ is constant as $n$ grows.

\emph{Notation}: We denote scalars by lower-case letters and sequences by capital letters. We denote the subsequence of $X$, from index $i$ to index $j$, with $i<j$ by $X(i : j)=x_ix_{i+1}\cdots x_j$. Matrices are denoted by boldfaced capitals. We use brackets for merging sequences, so $X=[X_1,\cdots,X_u]$ is a super-sequence defined by concatenating sequences $X_1,\cdots,X_u$.  Logarithms with base $2$ unless otherwise mentioned.

%%%%%%%%%%%%%%%%%%%%%%%%
%%%%%%%%%%%%%%%%%%%%%%%%

\section{Code Construction and Encoding} \label{sec:encoding}

We begin with a brief review of VT codes. For a detailed discussion on properties of VT codes the reader is referred to \cite{Sloane00}. The VT syndrome of a  binary sequence $W =(w_1,\ldots, w_n)$  is defined as 
\begin{equation}
\syn(W)=\sum_{j=1}^n j \,w_j \  \  (\text{mod }  (n+1)).
\end{equation}
For positive integers $n$ and $0\leq s\leq n$, we define the VT code of length $n$ and syndrome $s$, denoted by 
\be
\Vc\Tc_s(n) = \big\{W \in \{0,1\}^n: \syn(W)=s\big\} ,
\ee
as the set of sequences $W$ of length $n$ for which $\syn(W)=s$. %For example, 

The $(n+1)$ sets $\Vc\Tc_s(n) \subset \{0,1\}^n$, for $0\leq s\leq n$, partition the set of all sequences of length $n$.  Each of these sets $\Vc\Tc_s(n)$ is a single-deletion correcting code. The VT encoding and decoding complexity is linear in the code length $n$ \cite{AbdelFer98,Sloane00}. 

\subsection{Constructing the message $M$}\label{encodeing-sec}
The message  $M$ generated by the encoder consists of three parts, denoted by $M_1,M_2$, and $M_3$. The first part comprises the VT syndromes of  the blocks, the second part comprises the VT syndromes of the chunk-strings, and the third part is the parity-check syndrome of $X$ with respect to a linear code.

The sequence $X=x_1x_2\cdots x_n$ is divided into $l_1$ equal-sized blocks (assume that $n$ is divisible by $l_1$). We denote the length of each block  by $n_b=\frac{n}{l_1}$. For $1 \leq i \leq l_1$, the $i$th block is denoted by $B_i=X((i-1)n_b+1:in_b)$, and its VT syndrome is $s_{B_i}=\textsf{syn}(B_i)$. The first part of the message is the collection of VT syndromes for the $l_1$ blocks, i.e.,  $M_1= \{ s_{B_1}, s_{B_2},\cdots,s_{B_{l_1}} \}$. Since each $s_{B_i}$ is an integer between $0$ and $n_b$, the number of bits required to represent the VT syndromes of the $l_1$  blocks is $l_1 \lceil\log(n_b+1)\rceil$.

Each of the $l_1$ blocks is further divided into $l_2$ chunks, each of length $n_c$ bits. Since the length of each block is $n/l_1$, we have $n/l_1 = n_c l_2$, and therefore the length of $X$ satisfies  $n=n_c l_1 l_2$. (We assume that $\frac{n}{l_1}$ is divisible by $l_2$.)  For $1\leq j \leq l_2$, the $j$th chunk within the $i$th block is denoted by
\ben C^i_{j}=X((i-1)n_b+(j-1)n_c+1:(i-1)n_b+jn_c). \een
The $j$th \emph{chunk-string} is then formed by concatenating the $j$th chunk from each of the $l_1$ blocks. That is, the $j$th chunk string $C_j =[C^1_j,C^2_j,\cdots,C^{l_1}_j]$, for $1\leq j \leq l_2$. Fig. \ref{fig_enc} shows the blocks and the chunk-strings in an example  where $X$ of length $n=12$ is divided into $l_1=2$ blocks, each of which is divided into $l_2=2$ chunks of $n_c=3$ bits.  

The second part of the message is the collection of VT syndromes for the $l_2$ chunk-strings, i.e.,  $M_2= \{ s_{C_1}, s_{C_2},\cdots,s_{C_{l_2}} \}$, where $s_{C_j}$ denotes the VT syndrome of the $j$th chunk string.  Since the length of each chunk-string is $n_c l_1$, each $s_{C_j}$ is an integer between $0$ and $n_c l_1$. Therefore the number of bits required to represent the  VT syndromes of the $l_2$ chunk-strings is is $l_2 \lceil\log(n_c l_1+1)\rceil$.

The final part of the message is the  parity-check syndrome of $X$ with respect to a linear code. Consider a linear code of length $n$ with parity-check matrix $\boldsymbol{H}\in\{0,1\}^{z \times n}$. Then $M_3=\boldsymbol{H}X$ is the third component of $M$. The coset of the linear code containing $X$ will be used as an erasure correcting code. In our experiments in Section \ref{sec:simulation}, the linear code is chosen to be either a Reed-Solomon code over $GF(2^{n_c})$ or a bianry linear code defined by parity-check constraints drawn uniformly at random.   The number of bits for $M_3$ is equal to the number of rows of $\boldsymbol{H}$, i.e., number of binary parity checks in the code, 
$z$. If a  non-binary linear code with an $m \times n$ parity check matrix over $GF(2^{n_c})$ is used, the number of bits for $M_3$ 
is $z = m n_c$.

The total redundancy, or the overall number of bits required to represent the message $M=[M_1,M_2,M_3]$, is
$l_1 \lceil\log(n_b+1)\rceil+l_2 \lceil\log(n_cl_1+1)\rceil+ z$.

Since $n_b=n_cl_2$, normalizing by $n=n_c l_1l_2$ gives the \emph{synchronization rate}  (or per-symbol redundancy) $R_{\text{sync}}$ of our scheme:
\be R_{\text{sync}} = \frac{z}{n} + \frac{\lceil\log(n_cl_2+1)\rceil}{n_cl_2} + \frac{\lceil \log(n_cl_1+1) \rceil}{n_cl_1}.  \label{eq:sync_rate} \ee

\textbf{Remark.} To compute the per-symbol redundancy in \eqref{eq:sync_rate}, we assumed that each of the $(l_1+l_2)$ VT syndromes is separately converted to a binary sequence. The binary strings are then concatenated to construct the message. This can be done more efficiently: for instance, we can list all $(n_cl_2+1)^{l_1}(n_cl_1+1)^{l_2}$ possible syndromes, and use a look-up table to map these syndromes into binary sequences. This gives the following per-symbol redundancy:
\begin{align}
R_{\text{sync}} &= \frac{z}{n} + \frac{\lceil\log \left((n_cl_2+1)^{l_1}(n_cl_1+1)^{l_2}\right)\rceil}{n} \label{eq:sync_rate2}\\
&\leq \frac{z}{n} + \frac{\log(n_cl_2+1)}{n_cl_2} + \frac{\log(n_cl_1+1)}{n_cl_1}+\frac{1}{n}. \label{eq:sync_rate3}
\end{align} 
For the rest of the paper, unless specified, we use the expression in \eqref{eq:sync_rate} for the per-symbol redundancy.

To illustrate the effect of the code parameters on the redundancy, consider the following choice for synchronizing from $k$ deletions: $l_1 = l_2 = \alpha k$, for  $\alpha > 0$, so that  $n_c = n/(\alpha^2 k^2)$. Let the number of bits used to convey the parity check symbols be $z=\beta (k n_c)$, for $\beta \geq 0$.  With these parameters,  the per-symbol redundancy in  \eqref{eq:sync_rate} becomes 
\be
R_{\text{sync}} = \frac{\beta}{\alpha^2 k} +  \frac{2 \alpha k  \lceil \log(1 + n/(\alpha k)) \rceil}{n}.
\label{eq:rsync_special}
\ee
 The simulation results in Section \ref{sec:simulation} show that for $n$ in the range of a few hundreds to a few thousands, taking  $\alpha$ close to 1 gives a good tradeoff between redundancy, synchronization performance, and decoding complexity.     

Though we are primarily interested in constructing practical synchronization schemes for small and moderate values of $n$, it is interesting to examine how the redundancy scales with $n$  (with $k$ fixed). To achieve per-symbol redundancy of the optimal order $O( (k \log n)/n)$, we need to choose a constant $\alpha$ and set $\beta =0$, i.e., no parity check constraints. 
We discuss  this setting in Section \ref{sec:alt-dec}, where we use a guess-based decoder that  allows us to achieve the order-optimal redundancy at the expense of a slightly larger list size. Thus $\beta$ can be viewed as a tuning parameter that  allows us to tradeoff between list size and redundancy. In Section \ref{sec:conc}, we briefly discuss how the two-layer construction described above can be generalized to $L = \Theta(\log n)$ layers to achieve  
near-optimal redundancy even with the parity check constraints.

%%%%%%
\begin{example}\label{example_RS}
Suppose that we want to design a code for synchronizing a binary sequence of length $n=60$ from  $k=4$ deletions. Choose the chunk length $n_c=4$,  so that the sequence consists of 15 chunks.  Divide the sequence into $l_1=5$ blocks, each comprising $l_2=3$ chunks. Thus there are $5$ blocks each consisting of $3$ adjacent chunks, and $3$  chunk-strings each consisting of $5$ separated chunks. 

Noting that each chunk of $n_c=4$ bits corresponds to a symbol in $GF(2^4)$, we use a Reed-Solomon code defined over $GF(2^4)$ with length $2^4-1=15$. We also choose the parity-check matrix to have $4$ parity-check equations in $GF(2^4)$, so we can recover $4$ erased chunks using this Reed-Solomon code.

Assume that the sequence $X$ represented in $GF(2^4)$ is 
\be X=[ 4\ 10\ 5\quad 0\ 3\ 14\quad  7\ 7\ 1\quad 0\ 2\ 4\quad 4\ 6\ 8]^T. \ee
Each symbol above represents a chunk of $n_c=4$ bits. The first block $[4\ 10\ 5]$ in binary is
$B_1=0100\ 1010\ 0101.$
The VT syndrome of this sequence is $s_{B_1}=\syn(B_1)=10$.
The VT syndromes of the other four blocks are $6, 3, 4$, and $11$, respectively. The first part of the message is therefore $M_1=  \{ 10, 6, 3, 4 ,11 \}.$ 

We similarly compute $M_2$. The first chunk-string  $[4\ 0\ 7\ 0\ 4]$ in binary is
$$C_1=0100\ 0000\ 0111\ 0000\ 0100,$$
with VT syndrome $s_{C_1}=11$. Computing the VT syndromes of the other chunk-strings in a similar manner, we get
$M_2= \{ 11,20, 4 \}.$

The final part of the message is the syndrome of $X$ with respect to the Reed-Solomon parity-check matrix. We use the following parity-check matrix $\bsym{H}$ in $GF(2^4)$, with the generator $2$:
\ben
\bsym{H} =\begin{bmatrix}
1&1&1&1&\cdots &1 \\
1&2&4&8&\cdots&2^{14} \\
1&4&3&12&\cdots& (2^{2})^{14} \\ 
1&8&12&10&\cdots&(2^{3})^{14}\\
\end{bmatrix}
\een
to compute $M_3=\bsym{H}X = [11, 6, 13, 2]^T$. (In the representation of $GF(2^4)$ elements as degree-three polynomials with coefficients in $GF(2)$ with polynomial multiplication defined modulo $1+x +x^4$,  the generator $2$ corresponds to $x$.)
Since $\bsym{H}$ has four rows each representing one constraint in $GF(2^4)$, $z=16$ bits are needed to represent the parity-check syndrome in binary. The total number of bits to convey the message is 
$5\lceil\log(13)\rceil+3\lceil\log(21)\rceil+16=51 \text{ bits}.$
\end{example}

%%%%%%%%%%%%%%%%%%%%%%%%%%%%%%%%%%%
%%%%%%%%%%%%%%%%%%%%%%%%%%%%%%%%%%%

\section{Decoding Algorithm}\label{sec:decoding}

The goal of the decoder is to recover $X$ given $Y$, $n$ and the message $M=[M_1, M_2, M_3]$. 
From  $M_1,M_2$, the decoder knows the VT syndrome of each block and each chunk-string. Using this, the decoder first finds all possible configurations of deletions across blocks, and then for each of these configurations, it finds all possible chunk deletion patterns. Since each chunk is the intersection of a block and a chunk-string, each chunk plays a role in determining exactly two VT syndromes.  The intersecting construction of blocks and chunk-strings enables the decoder to iteratively recover the deletions in a large number of cases. The decoder is then able to localize the positions of the remaining deletions to within a few chunks. These chunks are considered erased, and are finally recovered by the erasure-correcting code.

The decoding algorithm consists of six steps, as described below.

%%%%

\begin{figure}[t]
\begin{center}
\begin{tikzpicture}[node distance=3.5cm,auto,>=latex', scale=1.4]
\draw[black, thick] (0,0) --(.5,1.5) node[font=\fontsize{9}{144}\selectfont,pos=1,anchor=south] {0};
\draw[black, thick] (0.5,1.5) -- (1,1.5); 
\draw[black, thick] (0,0) -- (0.5,0) node[font=\fontsize{9}{144}\selectfont,pos=1,anchor=south] {2};
\draw[black, thick] (0.5,0) -- (1,0); 
\draw[black, thick] (0,0)--(.5,-1) node[font=\fontsize{9}{144}\selectfont,pos=1,anchor=south] {3};
\draw[black, thick] (0.5,-1) -- (1,-1); 
\filldraw[black] (0,0) circle (1.5pt) node[font=\fontsize{9}{144},anchor=east] {a};   
    
\draw[black, thick](1,1.5)--(1.5,2.25)node[font=\fontsize{9}{144}\selectfont,pos=1,anchor=south] {1};
\draw[black, thick] (1.5,2.25) -- (2,2.25); 
\draw[black, thick](1,1.5)--(1.5,1.5)node[font=\fontsize{9}{144}\selectfont,pos=1,anchor=south] {2};
\draw[black, thick] (1.5,1.5) -- (2,1.5); 
\draw[black, thick](1,1.5) --(1.5,.75)node[font=\fontsize{9}{144}\selectfont,pos=1,anchor=south] {3};
\draw[black, thick] (1.5,.75) -- (2,.75);
\filldraw[black] (1,1.5) circle (1.5pt)node[font=\fontsize{9}{144},anchor=south] {b};   
    
\draw[black, thick](1,0)--(1.5,0)node[font=\fontsize{9}{144}\selectfont,pos=1,anchor=south] {1};
\draw[black, thick] (1.5,0) -- (2,0);
\filldraw[black] (1,0) circle (1.5pt)node[font=\fontsize{9}{144},anchor=south] {c};

\draw[black, thick] (1,-1) -- (2,-1);
\draw[black, thick] (1.35,-.85) -- (1.65,-1.15);
\draw[black, thick] (1.65,-.85) -- (1.35,-1.15);
\filldraw[black] (1,-1) circle (1.5pt)node[font=\fontsize{9}{144},anchor=south] {d};
\draw[black, thick,->] (2.25,-1) -- (2.75,-1) node[font=\fontsize{9}{144}\selectfont,pos=1,anchor=west] {Discarded};

\draw[black, thick] (2,0) -- (3,0);
\draw[black, thick] (2.35,.15) -- (2.65,-.15);
\draw[black, thick] (2.65,.15) -- (2.35,-.15);
\filldraw[black] (2,0) circle (1.5pt)node[font=\fontsize{9}{144},anchor=south] {h};
\draw[black, thick,->] (3.25,0) -- (3.75,0) node[font=\fontsize{9}{144}\selectfont,pos=1,anchor=west] {Discarded};

\draw[black, thick](2,2.25)--(2.5,2.25)node[font=\fontsize{9}{144}\selectfont,pos=1,anchor=south] {2};
\draw[black, thick] (2.5,2.25) -- (3,2.25);
\draw[black, thick] (2.35,.15) -- (2.65,-.15);
\draw[black, thick] (2.65,.15) -- (2.35,-.15);
\filldraw[black] (2,2.25) circle (1.5pt)node[font=\fontsize{9}{144},anchor=south] {e};
\draw[black, thick,->] (3.25,2.25) -- (3.75,2.25) node[font=\fontsize{9}{144}\selectfont,pos=1,anchor=west] {$(0,1,2)$};

\draw[black, thick](2,1.5)--(2.5,1.5)node[font=\fontsize{9}{144}\selectfont,pos=1,anchor=south] {1};
\draw[black, thick] (2.5,1.5) -- (3,1.5);
\draw[black, thick] (2.35,.15) -- (2.65,-.15);
\draw[black, thick] (2.65,.15) -- (2.35,-.15);
\filldraw[black] (2,1.5) circle (1.5pt)node[font=\fontsize{9}{144},anchor=south] {f};
\draw[black, thick,->] (3.25,1.5) -- (3.75,1.5) node[font=\fontsize{9}{144}\selectfont,pos=1,anchor=west] {$(0,2,1)$};

\draw[black, thick] (2,.75) -- (3,0.75);
\draw[black, thick] (2.35,.9) -- (2.65,.6);
\draw[black, thick] (2.65,.9) -- (2.35,0.6);
\filldraw[black] (2,.75) circle (1.5pt)node[font=\fontsize{9}{144},anchor=south] {g};
\draw[black, thick,->] (3.25,0.75) -- (3.75,0.75) node[font=\fontsize{9}{144}\selectfont,pos=1,anchor=west] {Discarded};

\filldraw[black] (3,2.25) circle (1.5pt)node[font=\fontsize{9}{144},anchor=south] {i};
\filldraw[black] (3,1.5) circle (1.5pt)node[font=\fontsize{9}{144},anchor=south] {j};
\end{tikzpicture}
\end{center}
\caption{Tree representing the valid block vectors for Example \ref{tree-ex}.} \label{Tree}
\vspace{-7pt}
\end{figure}

%%%%%
\subsection*{Step 1: Block boundaries}
In the first step, the decoder produces a list of candidate {\em block-deletion patterns} $V=(a_1,\cdots,a_{l_1})$ compatible with $Y$, where $a_i$ is the number of deletions in the $i$th block. Each pattern in the list should satisfy $\sum_{i=1}^{l_1} a_i=k$ with  $0\leq a_i\leq k$. The  list of candidates always includes the true block-deletion pattern. It is convenient to represent the candidate block-deletion patterns as branches on a tree with $l_1$ levels, as shown in Fig. \ref{Tree}. At every level (block) $i=1,\dotsc,l_1$, branches are added and labeled with all possible values of $a_i$. Specifically, the tree is constructed as follows.

\emph{Level $1$ of the tree}: Consider the first $n_b$ received bits $Y(1:n_b)$, compute its VT syndrome $u=\syn(Y(1:n_b))$ and compare it with $s_{B_1}$, the correct syndrome of the first block. There are two alternatives for the $k$ branches of the first level.

\begin{enumerate}
\item $\underline{u = s_{B_1}}$: 
First, the decoder adds a branch with $a_1=0$, corresponding to the case that the first $n_b$ bits are deletion-free. The first block cannot have just one deletion, because in this case the single-deletion correcting property of the VT code would imply that $u \neq s_{B_1}$.
However, it is possible that two or more than two deletions happened in block one, and by considering additional bits from the next block, the VT-syndrome of first $n_b$ bits accidentally matches with $s_{B_1}$. For example, consider blocks of length $n_b=4$, and let the first two blocks of $X$ be $ \underline{01}00 \ 1111\ \ldots$, 
with the underlined bits deleted we get $Y=001111 \ldots$. In this case $u=s_{B_1}=2$.
The decoder thus adds a branch for $a_1=0,2,\dotsc,k$.

\item $\underline{u \neq s_{B_1}}$: Block one contains one or more deletions and the decoder adds a branch for $a_1=1,2,\dotsc,k$.
\end{enumerate}

\emph{Level $i+1$, $1 \leq i  < l_1$}: Assume that we have constructed the tree up to level $i$. Consider a branch of the tree at level $i$ with  the number of deletions in blocks $1$ through $i$ given by   $a_1,a_2,\cdots,a_{i}$, respectively. This gives us the starting position of block $(i+1)$ in $Y$. Denote this starting position by 
\be p_{i+1}= n_bi-d_i+1.  \label{eq:pi1} \ee
where $d_i = \sum_{j=1}^{i}a_j$ is the number of deletions on the branch up to block $i$. Compute the VT syndrome of next $n_b$ bits  $u=\syn\left(Y\left( p_{i+1} : p_{i+1} +n_b-1\right)\right)$.
There are two alternatives:
\begin{enumerate}

\item $\underline{u = s_{B_{i+1}}}$: If $(k-d_i)<2$ then the only possibility is that $a_{i+1}=0$. If $(k-d_i)\geq 2$, $k-d_i-1$ branches are added for $a_{i+1} = 0,2,\dotsc,k-d_i$. 

\item $\underline{u \neq s_{B_{i+1}}}$: If $(k-d_i)>0$ then there are $(k-d_i)$ possibilities at this branch: the $i$th block can have $1, 2,\cdots, (k-d_i)$ deletions. If $(k-d_i)=0$, it is assumed this is an invalid branch, and the path is discarded.
\end{enumerate}

\begin{example}\label{tree-ex} Assume $k=3$ deletions, $l_1=3$ blocks, and that the true deletion pattern is (0,2,1), i.e., there are zero deletions in the first block,  two deletions in second block, and one deletion in third block.  The tree constructed by the decoder depends on the underlying sequences $X$ and $Y$. In Fig. \ref{Tree}, we illustrate one possible tree constructed for this scenario without explicitly specifying $X$ and $Y$.

 Assume that in the first step, the syndrome matches with $s_{B_1}$, so we have $a_1=0,2$, or $3$ . At node b (corresponding to $a_1=0$),  suppose that the syndrome does not match with $s_{B_2}$, so we have $a_2=1,2$, or 3.  Now suppose that at nodes c and d,  the syndrome does not match with  $s_{B_2}$. At node d,  $a_1=3$, so there are no more deletions available for the second block; so this branch is discarded. At node c, $a_1=2$, so the only possibility is one deletion in the second block. Then if the syndrome at node h  does not match $s_{B_3}$, the branch is discarded. At nodes e and f, we assign the remaining deletions to the last block. At node g, the syndrome does not match with $a_3$, and the branch is discarded.
\end{example}

\subsection*{Step 2: Primary fixing of blocks}

Denote by $\Lc_1$ the list of the block-deletion pattern candidates after the first step and denote the corresponding block-deletion patterns by $V_1,\cdots, V_{\vert\Lc_1\vert}$. 
In this second step, for each of the block-deletion patterns, we restore the deleted bit in blocks containing a single deletion by using the VT decoder. Specifically, for a block-deletion pattern $V=(a_1,\cdots,a_{l_1})$,  let the $i$th block of $Y$ with respect to $V$ be  $S= Y\left( p_i :  p_i +n_b-1\right)$
where $p_i$ is the starting position  of the $i$th block in $Y$, defined analogously to \eqref{eq:pi1}.
If $a_i=1$, feed the sequence $S$ to the VT decoder and in $Y$, replace  $S$ with the decoded sequence. After this, the $i$th block in $Y$ is deletion free, thus, the decoder updates the block-deletion pattern $V$ by setting $a_i=0$. We carry out this procedure for all blocks with one deletion in $V$. This results in  a sequence $\hat{Y}$, which is obtained from $Y$ by recovering the single-deletion blocks corresponding to block-deletion pattern $V$. Denote the updated version of block-deletion pattern $V$ by $\hat{V}$.
Thus at the end of this step, we have $\vert\Lc_1\vert$ updated candidate sequences 
$\hat{Y}_1,\cdots,\hat{Y}_{\vert\Lc_1\vert}$ with corresponding block-deletion patterns $\hat{V}_1,\cdots, \hat{V}_{\vert\Lc_1\vert}$.

%%%%
\begin{example} Consider the code of Example \ref{example_RS} with $l_1 = 5$ blocks, and $k=4$ deleted bits. If the list of block-deletion patterns at the end of the first step is
\begin{align*}
\begin{split}
V_1&=(1,1,1,1,0), \hspace{2mm} V_2=(1,1,2,0,0),\\
V_3&=(1,2,1,0,0), \hspace{2mm} V_4=(2,0,2,0,0),
\end{split} %\label{block-vector-2}
\end{align*}
then the updated list of block-deletion patterns is
\begin{align*}
\begin{split}
\hat{V}_1&=(0,0,0,0,0),  \hspace{2mm} \hat{V}_2=(0,0,2,0,0),\\
\hat{V}_3&=(0,2,0,0,0), \hspace{2mm} \hat{V}_4=(2,0,2,0,0).
\end{split} %\label{block-vector-2}
\end{align*}
\label{ex:Vtilde}
\end{example}
\subsection*{Step 3: Chunk Boundaries}\label{sec-chunk}
In this step, for each updated block-deletion pattern $\hat{V}$ and the corresponding $\hat{Y}$, we list all possible allocations of deletions across chunks. More precisely,  for each pair ($\hat{V}, \hat{Y}$) we list all possible $l_1\times l_2$ matrices  $\bm{A}=(a_{ij})$, where $a_{ij}$ is the number of deletions in the $j$th chunk of the $i$th block, such that  $\sum_{j=1}^{l_2} a_{ij} = a_i$, the $i$th entry of $\hat{V}$. The $j$th column of matrix $\bm{A}$, specifies the number of deletions in the $l_1$ chunks of the $j$th chunk-string. 
For example, some of the possible matrices for $\hat{V}_4=(2,0,2,0,0)$ in Example \ref{ex:Vtilde} are
\begin{align}
\begin{split}
\bm{A}_1=\begin{bmatrix}
1&1&0 \\
0&0&0 \\
0&1&1 \\
0&0&0 \\
0&0&0 
\end{bmatrix},
\
\bm{A}_2=\begin{bmatrix}
2&0&0 \\
0&0&0 \\
0&1&1 \\
0&0&0 \\
0&0&0 
\end{bmatrix},
\ 
\bm{A}_3=\begin{bmatrix}
1&0&1 \\
0&0&0 \\
1&0&1 \\
0&0&0 \\
0&0&0 
\end{bmatrix}.
\end{split} \label{matrices}
\end{align}

The algorithm that lists all chunk-deletion matrices $\bm{A}$ compatible with a given block-deletion pattern $\hat{V} =(a_1, \ldots, a_{l_1})$ is very similar to the tree construction described in Step 1. In this case, for each block-deletion pattern $\hat{V}$, another tree will be constructed, with each path in the tree representing a valid chunk-deletion matrix $\bm{A}$.

\emph{Level $1$ of the tree}: Construct a sequence $S$ by concatenating the first $n_c$ bits of each block in $\hat{Y}$ and compute its VT syndrome $u=\syn(S)$. There are two possibilities:
\begin{enumerate}
\item $\underline{u = s_{C_1}}$:  For the first chunk-string, list all valid chunk-deletion patterns of the form $(a_{11}, \ldots, a_{l_11})$, where $0\leq a_{i1} \leq a_i$, and $\sum_{i=1}^{l_1} a_{i1}\neq 1$, since a single deletion in the chunk-string would result in $u\neq s_{C_1}$. 

\item $\underline{u \neq s_{C_1}}$: List all valid chunk-vectors for the first chunk-string of the form $(a_{11}, \ldots, a_{l_11})$, where $0\leq a_{i1} \leq a_i$, and $\sum_{i=1}^{l_1} a_{i1}\geq 1$.
\end{enumerate}

\emph{Level $j$, $1 < j  \leq  l_2$}: Assume that we have constructed the tree up to level $(j-1)$. Thus, we know the number of deletions in each chunk of the first $(j-1)$ chunk-strings. From this, we can determine the total number of deletions in the first $(j-1)$ chunks of each block. Let $d_{i,j-1}$ denote the number of deletions in the first $(j-1)$ chunks of block $i$.   Then along this path,  the 
$j$th chunk of $i$th block in $\hat{Y}$ is
\begin{equation}
S_{ij}=\hat{Y}\Big(p_i+ (j-1)n_c-d_{i,j-1}:p_i + jn_c-d_{i,j-1}-1\Big).
\end{equation}
Form the $j$th chunk-string,
$S_{j}=[S_{1j},\cdots,S_{l_1j}]$, compute its VT syndrome $u=\syn(S_{j})$, and compare it with the correct syndrome $s_{C_{j}}$. There are two possibilities.
\begin{enumerate}
\item $\underline{u = s_{C_{j}}}$: List all valid chunk-deletion patterns for the $j$th chunk-string of the form $(a_{1j}, \ldots, a_{l_1j})$, where $0\leq a_{ij} \leq a_i-d_{i,j-1}$, and $\sum_{i=1}^{l_2} a_{ij}\neq 1$.

\item $\underline{u \neq s_{C_{j}}}$: List all valid chunk-deletion patterns for the $j$th chunk-string of the form $(a_{1j}, \ldots, a_{l_1j})$, where $0\leq a_{ij} \leq a_i-d_{i,j-1}$, and $\sum_{i=1}^{l_2} a_{ij}\geq 1$. 
If the list is empty, discard the branch. The list will be empty when there are no more deletions to assign to $j$th chunk-string. 
\end{enumerate}
At the end of step 3, the decoder provides a list of pairs $(\hat Y,\bm A)$, where $\hat Y$ is a candidate sequence to be decoded using the chunk-deletion matrix  $\bm A$, with $a_{ij}$ being the number of deletions in the $j$th chunk of the $i$th block. Denote the number of such pairs in the list by $\vert\Lc_3\vert$.

%%%%%%%%
\subsection*{Step 4: Iterative correction of blocks and chunk-strings}
Similar to step 2, in  step $4$ we use the  VT syndromes (known from the message sent by the encoder) to recover deletions in blocks and chunk-strings for which the matrix $\bm A$ indicates a single deletion. Whenever a  deletion recovered using a VT decoder lies in a chunk different from the one indicated by $\bm A$, the candidate is discarded. As discussed in Section \ref{sec:simulation} and \ref{sec:complexity}, this is an effective way of discarding several invalid candidates. The iterative algorithm is described below.  For each pair $(\hat Y,\bm{A})$:
\begin{enumerate}[i)]
\item For each column of $\bm{A}$ containing a single 1 (indicating a single deletion in the corresponding chunk-string), recover the deleted bit in the chunk-string using its VT syndrome. With some abuse of notation we still refer to the restored sequence as $\hat Y$. If the restored bit does not lie in the expected chunk indicated by the 1, discard the pair $(\hat{Y}, \bm A)$ and move to the next candidate pair. Otherwise, update the matrix $\bm{A}$ by replacing the 1s corresponding to the restored chunks by 0s. If there is a row in the updated matrix $\bm{A}$ with a single 1, proceed to step 4.ii). 

\item For each row of $\bm{A}$ containing a single 1 (indicating a single deletion in the corresponding block), recover the deleted bit in the block using its VT syndrome. Again, with some abuse of notation we still refer to the restored sequence as $\hat Y$. If the restored bit does not lie in the expected chunk indicated by the 1, discard the pair $(\hat{Y}, \bm A)$ and move to the next pair. Otherwise, update the chunk-deletion matrix by replacing the 1s corresponding to the restored chunks to 0s. If there is a column in the updated matrix $\bm{A}$ with a single 1, go to step 4.i).

\end{enumerate}

Denote the updated candidate pairs at the end of this procedure by $(\tilde Y,\tilde{\bm{A}})$, and assume there are $\vert\Lc_4\vert$ of them.

As an illustrative example, consider the three chunk-matrices given in \eqref{matrices}. In $\bm{A}_1$, we can successfully recover all the deletions. In 
$\bm{A}_2$, we can only fix two deletions in the third block. However, for $\bm{A}_3$, we cannot recover any of the deletions. Thus, the updated $\tilde{\bm A}$ matrices are
\begin{align}
\begin{split}
\tilde{\bm{A}}_1=\begin{bmatrix}
0&0&0 \\
0&0&0 \\
0&0&0 \\
0&0&0 \\
0&0&0 
\end{bmatrix},
\
\tilde{\bm{A}}_2=\begin{bmatrix}
2&0&0 \\
0&0&0 \\
0&0&0 \\
0&0&0 \\
0&0&0 
\end{bmatrix},
\ 
\tilde{\bm{A}}_3=\begin{bmatrix}
1&0&1 \\
0&0&0 \\
1&0&1 \\
0&0&0 \\
0&0&0 
\end{bmatrix}.
\end{split} \label{matrices1}
\end{align}
In Section \ref{sec:alt-dec}, we discuss a method to recover remaining deletions using VT constraints and bypassing the fifth step (where we use linear codes).

\subsection*{Step 5: Replacing deletions with erasures}\label{erasure-section}
In this step,  for each of the $\vert\Lc_4\vert$ surviving pairs $(\tilde Y, \tilde{\bm{A}} )$, we replace each chunk of $\tilde{Y}$ that still contains deletions with $n_c$ erasures. Hence, if there are $\nu$ chunks with deletions (where $1\leq\nu\leq k$), the resulting sequence will have length $n$, with $n_c \nu$ erasures and no deletions. Notice that this operation of replacing with erasures can be performed without ambiguity since $\tilde{\bm A}$  precisely indicates the starting position of each chunk and also the number of deletions within that chunk. 

The purpose of the linear code is to recover the erased bits.  The minimum distance of the linear code should be large enough to guarantee that we can resolve all the $\nu n_c$ erased bits. In Example \ref{example_RS}, as there are four deletions, we will have at most $\nu=4$ erased chunks, so we choose a Reed-Solomon code with $4$ parity-check equations in $GF(2^4)$. The chunk-matrix $\tilde{\bm{A}}_3$ in \eqref{matrices1} shows that a smaller number of  parity-check symbols will not suffice if we want to correct all deletion patterns. 

Some invalid candidates may be discarded  in the process of correcting the erasures as we may find that the parity-check equations are inconsistent, i.e. there is no solution for the erased chunks. We denote the number of remaining candidates at the end of this step by $\vert\Lc_5\vert$.

\subsection*{Step 6: Discarding invalid/identical candidates}
The reconstructed sequences at the end of Step $5$, denoted by $\hat X$,  all have length $n$ and are deletion free.  For each of the $\vert\Lc_5\vert$ sequences $\hat X$, we check the VT and parity-check constraints for each of the block and chunk-strings and discard those not meeting any of the constraints. 
At the end of Step 5 it is possible to have multiple copies of the same sequence. This is due to a deletion occurring in a run that intersects two chunks (or more); this deletion can be interpreted as a deletion in either chunk, and each interpretation leads to seemingly different candidates which will turn out to be the same at the end of the process. 
The surviving $\vert\Lc_6\vert$ distinct sequences comprise the final list produced by the decoder. 

The final list of reconstructed sequences consist of all length-$n$ sequences that can be obtained by adding $k$ bits to $Y$ and also satisfy all the VT and parity-check constraints. 
 The correct sequence is always among the $\vert\Lc_6\vert$ candidates. The synchronization algorithm is said to be zero-error if and only if $\vert\Lc_6\vert=1$ for all sequences and deletion patterns.  When $\vert\Lc_6\vert >1$, the list size can be further reduced if  additional hash functions or cyclic redundancy checks are available from the encoder.

%%%%%%%%

\section{Numerical experiments} \label{sec:simulation}

In this section, we present numerical results illustrating the performance of the synchronization code for various choices of the system parameters. The different setups that were simulated are shown in Table \ref{Set-up-table}.  For each setup, the performance was recorded over $10^6$ trials. In each trial, the sequence $X$ and the locations of the $k$ deletions were chosen independently and uniformly at random. For the first five setups, we used parity-check constraints from a Reed-Solomon code over $GF(2^{n_c})$ with code length $(2^{n_c}-1)$. For example, in setup $5$ we used $7$  parity-check constraints from a Reed-Solomon code over $GF(2^{6})$, hence $z=42$ bits are needed to represent the parity-check syndrome. In the last three setups, where the $z$ entry is denoted with an asterisk, we used $z$ binary parity-check constraints (for a length $n$ seqeunce) drawn uniformly at random. 

\begin{table}[t!]
\centering
\caption{Number of deletions $k$, code length $n$, and code parameters for each setup. \vspace{1mm}}\label{Set-up-table}
\begin{tabular}{cccccccc} 
\toprule
  &  $k$ &$n$& $l_1$ & $l_2$ & $n_c$ & $z$  &  $R_{\text{sync}}$\\
\midrule   
Setup 1 &3 & 60  &   5 & 3 & 4 & 4& 0.650\\
Setup 2 & 3 & 60  &    5 & 3 & 4 & 8 & 0.717\\
Setup 3 & 3 &60  & 5 & 3 & 4 & 12 & 0.783\\
Setup 4 & 4  &60  &  5 & 3 & 4 & 16 & 0.850\\
Setup 5 & 7 & 378  &  9 & 7 & 6 & 42 & 0.365\\
Setup 6 & 7& 486  & 9 & 9 & 6 & $50^*$ & 0.325\\
Setup 7 & 9& 1080& 15& 12 & 6& 55*& 0.225   \\
Setup 8 & 10& 2800 &  20& 20& 7 & $60^*$ & 0.135 \\ 
\bottomrule 
\end{tabular}
\bigskip
%\end{table}
%%%%%%
%\begin{table}[ht]
\centering
\caption{List size after each step, averaged over $10^6$ trials.\vspace{1mm}}\label{Deletion-table}
\resizebox{\columnwidth}{!}{
\begin{tabular}{cccccccc} 
\toprule
&$\mathbb{E}\vert\Lc_1\vert$& $\mathbb{E}\vert\Lc_3\vert$ & $\mathbb{E}\vert\Lc_4\vert$  & $\mathbb{E}\vert\Lc_6\vert$ &  $\max \vert\Lc_6\vert$&  $\mathbb{P}[\vert\Lc_6\vert >1]$ \\
\midrule   
Setup 1 	  & 1.87 & 1.92	& 1.42 & 1.003 & 3 &0.003  \\
Setup 2 	  & 1.87 & 1.92 & 1.42 & 1.000 &2 &$2.5\times 10^{-5}$  \\
Setup 3 	  & 1.87 & 1.92 & 1.42 & 1      &1 &0  \\
Setup 4      & 3.39 & 6.18 & 2.53 & 1 	   &1 &0  \\
Setup 5      & 11.51 & 74.43& 3.42& 1      & 1&0   \\
Setup 6      & 11.20 & 28.64& 2.55& 1      & 1&0  \\
Setup 7    & 14.45 & 94.38& 2.41& 1      & 1&0  \\
Setup 8      & 12.76 & 26.16& 1.57& 1      & 1&0  \\
\bottomrule 
\end{tabular}
}
\vspace{-5pt}
\end{table}
%%%%%%%

Table \ref{Deletion-table} shows the  list sizes of the number of candidates at the end of various steps of the decoding process. Recall that $\vert\Lc_1\vert$ is the number of candidate block-deletion patterns at the end of step $1$, $\vert\Lc_3\vert$  is the number of pairs $(\hat Y,\bm A)$ at the end of step $3$,  $\vert\Lc_4\vert$  is the number of pairs $(\tilde Y, \tilde{\bm A})$ at the end of step $4$, and $\vert\Lc_6\vert$ is the number of sequences $\hat X$ in the final list. The average of $\vert\Lc_i\vert$ over the $10^6$ trials is denoted by $\mathbb{E}\vert\Lc_i\vert$. The column $\max \vert\Lc_6\vert$ shows the maximum size of the final list across the $10^6$ trials. The column $\mathbb{P}[\vert\Lc_6\vert >1]$ shows the fraction of trials for which $\vert\Lc_6\vert>1$.

The first three setups have identical parameters, except for the number of Reed-Solomon  parity checks. This shows the effect of adding parity-check constraints on the list size and the redundancy. Adding more parity-check constraints improves the decoder performance by reducing the number of trials with list size greater than one, at the expense of increased redundancy.

The fourth setup is precisely the code described in Example \ref{example_RS}. It has the same values of $(n_c,l_1, l_2)$ as the first three setups but with a larger number of deletions and parity-check constraints. We observe that increasing the number of deletions  (with $n_c,l_1, l_2$ unchanged)  increases the average number of candidates in the different decoding steps. In general, choosing $l_1\geq k$ ensures that the average list size after step $1$ is small. 

The fifth setup is a larger code with length $n=378$ and can handle a larger number of deletions ($k=7$). Though the final list size is always one, the number of candidate chunk-deletion matrices at the end of the third step is large, which increases the decoding complexity. The only difference between setups five and six is that the latter has a larger value of $l_2$. Comparing $\mathbb{E}\vert\Lc_3\vert$ for these setups, we observe that increasing $l_2$ significantly reduces the number of candidate chunk-deletion matrices at the end of the third step. This is because increasing $l_2$ increases the number of chunk-string VT constraints, which allows the decoder to eliminate more candidates while determining chunk boundaries. 

The last setup is a relatively long code. Although the average number of candidates in each of the decoding steps is not very high, we found that a small fraction of trials have a very large number of candidates, resulting in considerably slower decoding for these trials.

\section{List size analysis}
\label{sec:listsize}

The final list produced by the decoder consists of all sequences that satisfy the $l_1$ block VT constraints, the $l_2$ chunk-string VT constraints, and the  parity-check constraints.  Recall that at the end of step 3 of decoding we have a set of candidate chunk deletion patterns, each of which is of the form $\{ a_{ij} \}_{1\leq i \leq l_1, \, 1 \leq j \leq l_2}$, where $a_{ij}$ is the number of deletions in chunk $j$ within block $i$. A number of candidate patterns are then discarded in Step 4 as they fail to satisfy the intersecting VT constraints.   

As evident from Table \ref{Deletion-table}, the VT constraints play a key role in reducing the list size. However, the non-linearity of the VT constraints and the intersecting construction makes it challenging to obtain theoretical bounds on the list size. We will therefore bound the expected list size by considering only the effect of the parity-check constraints. Though the bound is loose, it gives us insight into how the code parameters affect the list size.

In step 5 of decoding, the parity-check constraints are used to recover the unresolved chunks for each of the surviving chunk deletion patterns at the end of step 4.   Since there are a total of $k$ deletions, we consider all chunk-deletion patterns
$\{ a_{ij} \}_{1\leq i \leq l_1, \, 1 \leq j \leq l_2}$ that satisfy
\be
\sum_{i=1}^{l_1} \sum_{j=1}^{l_2} a_{ij} =k, \quad a_{ij} \geq 0. 
\label{eq:sum_deletions}
\ee
Furthermore, assume that any pattern of upto $k$ erased chunks can be recovered using the $z$ binary parity-check constraints.  This can be ensured by using $z = k n_c$ linearly independent parity-check equations from a binary linear code with minimum distance at least $k n_c +1$. (For example, we can use $k$ parity-check constraints of an $(n-k, n)$ MDS code over GF($2^{n_c})$.)
 This implies that the parity-check constraints can be used to recover any pattern of up to $k$ erased chunks.   For each chunk-deletion pattern considered, the bits in the unresolved chunks are erased (according to the pattern), and the parity-check constraints are used to recover these erased chunks.   Note that the recovered bits should be a supersequence of the bits erased in the unresolved chunks, otherwise the decoder can discard the deletion pattern. 

We will  bound the  the probability that an incorrect deletion pattern satisfying \eqref{eq:sum_deletions} satisfies all the parity-check constraints and is a supersequence of the erased bits. Since there are $\binom{k+l_1l_2-1}{k}$ deletion patterns satisfying \eqref{eq:sum_deletions}, this will give a bound on the expected list size.  We make two assumptions on any sequence reconstructed using an incorrect chunk deletion pattern.  To motivate these assumptions, consider the following example.

\begin{example}
Assume that $n_c=3$ and $l_1=l_2=2$, and that are $k=3$ deletions. Let $X=\underline{1}01\ 100\ 011\ 1\underline{00}$ and $Y=011 000 111$, with the underlined bits being deleted from $X$ to produce $Y$. The correct chunk deletion pattern is $(1, 0, 0, 2)$. According to this pattern, the erased sequence $Y'= xxx \ 100 \ 011\ xxx$ where $x$ denotes an erased bit (from which $X$ is recovered using the parity-check constraints. 

 Consider an incorrect deletion pattern, say $(2,0, 0, 1)$. The erased sequence according to this pattern is $Y''=xxx \ 110\ 001\ xxx$. Note that for recovered sequence based on this deletion pattern, the decoder requires  that  the first chunk contains a 0, and the last chunk contains 11. We will assume that the bits recovered in first and fourth chunks of $Y''$ (using the parity-check constraints) are uniformly random and hence independent of the erased bits ($0$ in the first chunk and $11$ in the fourth chunk). 
\end{example}

\textbf{Assumption 1.} If the set of parity-check constraints has a solution for the erased chunks corresponding to an incorrect deletion pattern, then the recovered bits will be uniformly random and independent of the bits erased from the chunks.  

Given an incorrect deletion pattern which corresponds to $k' < k$ erased chunks, the chunks can be recovered using any $k'$ parity-check constraints of the MDS code. 

\textbf{Assumption 2.} Given an incorrect deletion pattern for which the erased chunks are recovered using $z' < z$ binary parity-check constraints  then the recovered sequence satisfies each of the remaining $(z-z')$ constraints independently with probability $\frac{1}{2}$. Therefore the probability that the recovered sequence satisfies all the remaining parity-check constraints is $\left(\frac{1}{2}\right)^{z-z'}$.

Assumption 2 is motivated by the observation that when for an incorrect  chunk deletion pattern, the bits in the $i$th unerased chunk in the sequence do not represent the actual $i$th chunk of the sequence.  Furthermore, using Assumption 1, the recovered bits in the erased chunks are uniformly random. Hence  evaluating a new parity-check constraint on the recovered sequence is equally likely to result in 0 or 1.
\begin{proposition}
Assume that the binary string $X$ and the locations of the $k$ deletions to produce $Y$ are both chosen uniformly at random. 
Let the $z \geq k n_c$  linearly independent binary parity-check constraints be chosen from a linear code with minimum distance at least $k n_c+1$ bits.  
Then under Assumptions 1 and 2, the probability that the final list size exceeds 1 satisfies
\be 
\mathbb{P}[\vert\Lc_6\vert >1]\leq \left( e \left( 1 + \frac{l_1 l_2}{k} \right) \left( \frac{n_c+1}{2^{n_c}} \right) \right)^k.
\ee 
The expected size of the final list satisfies
\be 
\mathbb{E}\vert \Lc_6\vert \leq 1+\left( e \left( 1 + \frac{l_1 l_2}{k} \right) \left( \frac{n_c+1}{2^{n_c}} \right) \right)^k.
\label{eq:expec_list_size}
\ee
\label{prop:list_size}
\end{proposition}
\begin{IEEEproof}
Consider an incorrect deletion pattern $(a_{11},\cdots,a_{l_1l_2})$, where the deletions are in $k'\leq k$ chunks. Let $a_{ij} >0$ in this pattern. Using Assumption 1, the probability that the $n_c$ bits recovered  in this chunk (using the parity-check constraints) are a supersequence of the $(n_c-a_{ij})$ bits erased in this chunk is:
\be\label{eq:super:guess}
\frac{\sum_{m=0}^{a_{ij}}\binom{n_c}{m}}{2^{n_c}}\leq \frac{(n_c+1)^{a_{ij}}}{2^{n_c}}.
\ee
The numerator in the LHS is the number of supersequences of length $n_c$ for the  erased chunk (see \cite{levenshtein2001} for a proof), and the denominator is the total number of length $n_c$ binary sequences. Therefore, the probability that the recovered  sequence is a supersequence of the erased bits in all the chunks with deletions can be bounded by
\be
\label{eq:super_bound2}
\prod_{i,j:a_{ij}\geq1}\left(\frac{(n_c+1)^{a_{ij}}}{2^{n_c}}\right) = \frac{(n_c+1)^k}{2^{k'n_c}},
\ee
where we have used  $\Sigma_{i,j} a_{i,j} =k$ and the fact that the deletion pattern has $k'$ chunks with deletions, i.e., $k'$ pairs $(i,j)$ with $a_{i,j} >0$.

Since $k'$ chunks are erased, $k'n_c$ linearly independent parity constraints suffice to recover the $k'n_c$ bits in these chunks.  Furthermore, using Assumption 2, the probability that the recovered sequence satisfies the  $(z-k'n_c)$ remaining parity-check equations is  $\big(\frac{1}{2}\big)^{z-k'n_c} \leq \big(\frac{1}{2}\big)^{kn_c-k'n_c}$.  Combining this with \eqref{eq:super_bound2}, we have the following upper bound on the probability that the sequence recovered from the  incorrect chunk deletion pattern is  in the final list:
\be
\left( \frac{1}{2} \right)^{kn_c -k' n_c} \frac{(n_c+1)^k}{2^{k'n_c}}  =\left(\frac{n_c+1}{2^{n_c}}\right)^k. \label{eq:upper:size}
\ee
Now using union bound for each of the possible chunk deletion patterns, the probability that the final list size exceeds 1 satisfies
\begin{align}
\mathbb{P}[\vert\Lc_6\vert >1]&\leq \binom{k+l_1l_2-1}{k}\left(\frac{n_c+1}{2^{n_c}}\right)^k\\
 &= \left( e \left( 1 + \frac{l_1 l_2}{k} \right) \left( \frac{n_c+1}{2^{n_c}} \right) \right)^k,
\end{align}  

We can also use the upper bound in \eqref{eq:upper:size} for each of the possible chunk deletion patterns. Noting that the correct pattern will be on the list with probability 1, we have

\begin{align}
\mathbb{E}\vert \Lc_6\vert &\leq 1+ \binom{k+l_1l_2-1}{k}\left(\frac{n_c+1}{2^{n_c}}\right)^k\\
&\leq 1+\left( \frac{e(k+l_1l_2)}{k}\right)^k\left(\frac{n_c+1}{2^{n_c}}\right)^k.
\end{align}
\end{IEEEproof}

According to Proposition \ref{prop:list_size}, we will have $\mathbb{E}\vert \Lc_6\vert<2$ if the parameters are chosen such that
\be 
n_c\log 2 - \log(n_c+1)>\log e + \log (1+l_1l_2/k) .
\ee
For example, we can choose $l_1 = l_2 =k$, and $n _c > \log (3 (1 +k))$ to ensure that $\mathbb{E}\vert \Lc_6\vert<2$ as the number of deletions $k$ grows. This choice is similar to the parameters used for the numerical simulations in Section \ref{sec:simulation}.

Since we have not taken into account the effect of the VT constraints,  the bound in \eqref{eq:expec_list_size}  is loose. Indeed, the bound suggests that increasing $l_1, l_2$ will increase the final list size. However, we will see in the next section that  increasing $l_1, l_2$  reduces the average list size in the intermediate steps of the decoder. This is because increasing the number of VT constraints allows the decoder to  reject a larger of number of incorrect deletion patterns as being inconsistent with the VT constraints. An interesting direction for future work is to  tighten the bound of  Proposition \ref{prop:list_size} by taking into account the effect of the VT constraints.

%%%%%%%%%%%%%%%%%%%%%%%%%%%%%%%%%%%
%%%%%%%%%%%%%%%%%%%%%%%%%%%%%%%%%%%
 
 \section{Encoding and decoding complexity}
 \label{sec:complexity}
 
 In this section we discuss the number of operations required for constructing the encoded message $M=[M_1,M_2,M_3]$ and for the decoding algorithm.
 
\subsection{Encoding complexity}\label{sec:enc_complexity}
 Computing the VT syndrome of a length $m$ sequence needs $O(m)$ arithmetic operations. For  $M_1$ we need to compute VT syndrome of $l_1$ blocks, each of length $n_b=n_c l_2$. This can be done with $O(n_c l_1  l_2)$ operations. Similarly, for $M_2$ we need $ O(n_c l_2 l_1)$  operations. Recalling that $n=n_cl_1l_2$,  the complexity of computing $M_1$ and $M_2$ is $O(n)$. Recall that $M_3=\boldsymbol{H}X$ is constructed via multiplication of a $z\times n$ matrix with a length $n$ vector, which requires $O(zn)$ operations. This is the dominant term in the encoding complexity, therefore, the overall complexity of the encoder is $O(zn)$.

\subsection{Decoding Complexity}\label{sec:dec_complexity}
In the following, we analyze the decoding complexity by finding an upper bound for the complexity of each of the six decoding steps.
\subsection*{Step 1: Block boundaries}
In the first step, we construct a tree for finding all the candidates for block boundaries. At each node of the tree, a VT syndrome of a length $n_b$ sequence is computed and compared with the syndrome known from $M_1$.  
Since there are a total of $l_1$ blocks, the maximum number of valid block deletion patterns at the end of Step $1$ is the number of  non-negative integer solutions of
\be 
a_1+\cdots+ a_{l_1}=k, 
\label{block-eq}
\ee  
which is $\binom{k+l_1-1}{k}$. This number is only an upper bound on the number of valid block deletion patterns as we do not take into account  the effect of the VT syndrome in discarding patterns. (Recall that for a block deletion pattern to be valid, all the blocks with zero deletions in the pattern should be consistent with their VT syndromes.) Each valid block deletion pattern is a leaf of the tree. Since there are  $l_1$ levels, the total number of nodes in the tree is at most $l_1\binom{k+l_1-1}{k}$. (Note from Fig. \ref{Tree} that not every block deletion pattern corresponds to a branch with $l_1$ nodes.)

Therefore, an upper bound for the number of required operations in step 1 is:
\be \binom{k+l_1-1}{k}\times l_1 \times O(n_b)= O\left(n\binom{k+l_1-1}{k}\right).\ee

As explained in the Section \ref{sec:decoding}, many of the branches of the tree will get discarded because of the block deletion pattern being inconsistent with the VT constraints. The average number of nodes at the final level of the tree (denoted by $\mathbb{E}\vert\Lc_1\vert$) is particularly important since it will determine the average complexity of the next steps of the decoding. In Table \ref{compare-table}, we compare the empirical value of $\mathbb{E}\vert\Lc_1\vert$ (from Table \ref{Deletion-table}) with $\binom{k+l_1-1}{k}$, the upper bound for $\vert\Lc_1\vert$ obtained from \eqref{block-eq}. The considerable difference between these two numbers shows the importance of using VT codes --- in addition to recovering single deletions, they  act as hashes and allow the decoder to discard a larger number of incorrect block deletion patterns. This significantly decreases the decoding complexity by reducing the number of candidates that need to be considered in subsequent steps. This lower complexity allows us to efficiently decode relatively long codes like the code in setup 7 of Table \ref{Set-up-table}.

\begin{table}[t]
\centering
\caption{Comparison of average number of surviving paths after Step 1.\vspace{1mm}}
\resizebox{\columnwidth}{!}{
\begin{tabular}{cccccccc} 
\toprule
&Setup 1 to 3& Setup 4 & Setup 5 & Setup 6 & Setup 7   \\
\midrule   
$\mathbb{E}\vert\Lc_1\vert$       & $1.87$ & $3.39$ & $11.51$ & $11.20$ & $12.76$    \\
$\binom{k+l_1-1}{k}$        & $35$   & $70$   & $6.4\times 10^3$  & $6.4\times 10^3$  & $2.0\times 10^7$ \\
\bottomrule 
\end{tabular}
}
\label{compare-table}
\end{table}
In Figures \ref{fig:2figsA} and \ref{fig:2figsB}, we show how the empirical average $\mathbb{E}\vert\Lc_1\vert$ changes with the parameters of the code. The  list size $\abs{\Lc_1}$ depends on the number of deletions $k$, the number of block constraints $l_1$ and the number of  bits per block $n_b$. %(recall that $n_b=l_2n_c$). 
In Figure \ref{fig:2figsA}, where $k$ and $n_b$ are fixed, we see that  the empirical average $\mathbb{E}\vert\Lc_1\vert$ decreases with $l_1$ (although $\binom{k+l_1-1}{k}$, the upper bound on $\vert\Lc_1\vert$ increases). This is because a given solution of equation \eqref{block-eq} will not be in the list  $\Lc_1$ when it is not consistent with a block VT constraint. In particular, if $a_i=0$, i.e., the block is considered deletion free according to the deletion pattern,  the VT syndrome of the sequence corresponding to $i$th block should match with the correct syndrome known from $M_1$.

Figure \ref{fig:2figsB} shows that $\mathbb{E}\vert\Lc_1\vert$ also decreases with $n_b$ when $k$ and $l_1$ are fixed. This is because the probability that the VT syndrome of a length $n_b$ sequence accidentally matches  the correct block VT syndrome decreases with $n_b$. (Recall that that the VT syndrome is a number between $0$ and $n_b$.)  Such accidental matches, if  not detected in a subsequent level of the tree,  will increase the number of  incorrect deletion patterns in  $\Lc_1$.

\begin{figure}[t]
\begin{minipage}{.5\textwidth}
\begin{tikzpicture}[scale=0.6]
\begin{axis}[%
width=4.521in,
height=3.566in,
at={(0.758in,0.481in)},
scale only axis,
xlabel={\large Number of blocks $l_1$},
ylabel={\large $\mathbb{E}\vert\Lc_1\vert$},
xmin=5,
xmax=20,
ymin=5,
ymax=50,
axis background/.style={fill=white}
]
\addplot [color=black,line width=2.0pt, forget plot]
  table[row sep=crcr]{%
5    47.362462\\
6    45.116729\\
7    37.476943\\
8    29.74525\\
9    23.562396\\
10    18.95927\\
11    15.556039\\
12    13.035612\\
13    11.120063\\
14    9.638394\\
15    8.460148\\
16    7.536662\\
17    6.777879\\
18    6.146297\\
19    5.640274\\
20    5.199552\\
};
\end{axis}
\end{tikzpicture}
\caption{Empirical average $\mathbb{E}\vert\Lc_1\vert$ for different values of $l_1$\\ when $k=8$,  and block length $n_b =42$.}
\vspace{10pt}
\label{fig:2figsA} 
\end{minipage}
\begin{minipage}{.5\textwidth}
\begin{tikzpicture}[scale=0.6]
\begin{axis}[%
width=4.521in,
height=3.566in,
at={(0.758in,0.481in)},
scale only axis,
xlabel={\large Block length $n_b$},
ylabel={\large $\mathbb{E}\vert\Lc_1\vert$},
xmin=21,
xmax=105,
ymin=22,
ymax=26,
axis background/.style={fill=white}
]
\addplot [color=black,line width=2.0pt, forget plot]
  table[row sep=crcr]{%
21    25.978634\\
28    24.762449\\
35    24.040461\\
42    23.564042\\
49    23.190008\\
56    22.937543\\
63    22.754566\\
70    22.609251\\
77    22.459335\\
84    22.33285\\
91    22.274028\\
98    22.215372\\
105    22.119563\\
};
\end{axis}
\end{tikzpicture}
\caption{Empirical average $\mathbb{E}\vert\Lc_1\vert$ for different values of $n_b$ when $k=8$, and the number of blocks $l_1=9$.} %, and $l_2=7$.}
\label{fig:2figsB}
\end{minipage}
\end{figure}

\subsection*{Step 2: Primary fixing of blocks} 

In the second step, we use block VT syndromes to recover deletions in blocks with a single deletion. There are at most $l_1$ such blocks. Since the VT decoding complexity is linear in $n_b$ (the length of each block), the complexity for the second step is 
\be \vert\Lc_1\vert \times l_1\times O(n_b)=O\left(n\vert\Lc_1\vert\right).\ee

\subsection*{Step 3: Chunk Boundaries} 

In this step, we find all possibilities for the number of deletions in each chunk by performing the tree search on each of the block deletion patterns produced in the first step. Let $V=(a_1,\cdots,a_{l_1})$ to be one of the block deletion patterns at the end of the first step. Without loss of generality, assume that $a_1, a_2, \ldots, a_s$ are non-zero, for some  $s \leq l_1$. Since these $s$ blocks are not recovered in the second step of the decoding we know that $a_1, \ldots, a_s$ are each greater than 1. Furthermore, $\sum_{i=1}^s a_i\leq k$. Recalling that $a_{ij}$ represents the number of deletions in the $j$th chunk of the $i$th block, we have
\begin{align}\label{system-eqs}
\begin{split}
a_{11}+a_{12}+\cdots+a_{1l_2}&=a_1\\
a_{21}+a_{22}+\cdots+a_{2l_2}&=a_2\\
&\vdots\\
a_{sl_2}+a_{sl_2}+\cdots+a_{sl_2}&=a_s.\\
\end{split}
\end{align}
Similar to the first step, the number of non-negative integer solutions of the above set of equations is an upper bound for the number of nodes in the last level of the tree which can also serve as an upper bound for the other levels. To bound the complexity of this step we need the following lemma.
\begin{lemma}\label{eq-lemma}
The number of non-negative integer solutions of the set of equations in \eqref{system-eqs} when 
$\sum_{i=1}^s a_i=k$, $a_i\geq 0$, and $s$ and $l_2$ are positive integers, is  bounded by 
\be \binom{k/s+l_2-1}{l_2-1}^s. \label{step3-upper} \ee
Here, for a real number $x$ and integer $a$,
\be 
\label{def-binom}  \binom{x}{a}\triangleq \frac{x(x-1)\cdots (x-a+1)}{a!}. 
\ee
\end{lemma}
\begin{proof}
See Appendix \ref{App-lemma1}.
\end{proof}
Lemma \ref{eq-lemma} shows that \eqref{step3-upper} is an upper bound for the number of nodes in each level of the tree corresponding to the block deletion pattern $(a_1,\cdots,a_{l_1})$. Since $\sum_{i=1}^{l_1} a_i=k$ and $a_i\geq 2$ for $1 \leq i \leq s$,  $s$ is a number between $1$ and $\frac{k}{2}$. It is shown in Appendix \ref{app:der} that the derivative of \eqref{step3-upper} with respect to $s$ is positive when $s>1$. Therefore, $s=\frac{k}{2}$ maximizes \eqref{step3-upper}. Thus an upper bound for the number of nodes in each level of the tree is 
\be \max_{1\leq s\leq \frac{k}{2}} \binom{k/s+l_2-1}{l_2-1}^s=\binom{l_2+1}{l_2-1}^{\frac{k}{2}}=\binom{l_2+1}{2}^{\frac{k}{2}}. \label{eq:max:step3}\ee
We therefore have  
\be \vert\Lc_3\vert\leq  \vert\Lc_1\vert \binom{l_2+1}{2}^{\frac{k}{2}}. \label{eq:up:step3}\ee
At each node of the tree, we compute the VT syndrome of a length $n_cl_1$ sequence and compare it with the syndrome known from $M_2$. Therefore, the complexity of this step is 
\begin{align}
\vert\Lc_3\vert \times l_2 \times O(n_cl_1) \leq O\left(n\vert\Lc_1\vert \binom{l_2+1}{2}^{\frac{k}{2}}\right)
&=O\left(n\vert\Lc_1\vert l_2^k\right).
\end{align}  
Similar to the first step, many of the solutions of \eqref{system-eqs} are not compatible with VT syndromes of chunk-strings. Table \ref{compare-table2} compares the empirical value of $\mathbb{E}\vert\Lc_3\vert$ with the upper bound in \eqref{eq:up:step3}, and shows the importance of  the  VT constraints in reducing the  number of compatible chunk deletion patterns in Step 3. 

\begin{table}[t]
\centering
\caption{Comparison of average number of surviving paths after Step 3.\vspace{1mm}}
\resizebox{\columnwidth}{!}{
\begin{tabular}{cccccccc} 
\toprule
&Setup 1 to 3& Setup 4 & Setup 5 & Setup 6 & Setup 7   \\
\midrule   
$\mathbb{E}\vert\Lc_3\vert$       & $1.92$ & $6.18$ & $74.43$ & $28.64$ & $26.16$    \\
$\mathbb{E}\vert\Lc_1\vert \binom{l_2+1}{2}^{k/2}$        & $27.48$   & $122.04$  & $1.3\times 10^6$  & $6.8\times 10^6$  & $5.2\times 10^{12}$ \\
\bottomrule 
\end{tabular}
}
\label{compare-table2}
\vspace{-4pt}
\end{table}

\subsection*{Step 4: Iterative correction of blocks and chunk-strings}

In this step, we iteratively use the VT decoder for blocks and chunk strings to recover deletions. Each of the VT checks will be used at most once. Since there are $l_1$ blocks and $l_2$ chunk-strings an upper bound for the complexity is
\be \vert\Lc_3\vert \times \left(l_1\times O(n_cl_2)+l_2\times O(n_cl_1)\right)= O\left(n\vert\Lc_3\vert\right). \ee
Recall from the decoding algorithm that some of the candidates will be discarded in this step, therefore, $\vert\Lc_4\vert\leq \vert\Lc_3\vert$. 
\subsection*{Step 5: Replacing deletions with erasures}
In this step, we use the linear equations for recovering the erased chunks. There are at most $k$ erased chunks and hence $kn_c$ bits erased. Hence, the complexity of finding solutions for the set of linear equations can be  bounded by
$O\left(n^3\vert\Lc_4\vert\right)$.
We discard a candidate if there is no solution for the linear equations; therefore $\vert\Lc_5\vert\leq \vert\Lc_4\vert$.

\subsection*{Step 6: Discarding invalid/identical candidates} 

In this step, we compute the VT syndrome of blocks and chunk-strings for all the candidates on the list and compare them with the known syndromes. Hence the complexity is $O\left(n\vert\Lc_5\vert\right)$.

We have computed the complexity of each step of the decoding in terms of the the list size at the end of the previous step. An upper bound for the decoding complexity (not considering the effect of VT codes in eliminating incompatible deletion patterns) is $O\left(n^3\binom{k+l_1-1}{k}l_2^k\right)$. 
If one assumes that $k$, $l_2$, and $l_1$ are fixed and the length of the code is increased by increasing $n_c$, then the complexity of the decoding is $O(n^3)$ while the complexity of the encoding is $O(n)$. We remark again that this bound on decoding complexity is loose: as illustrated in Tables \ref{compare-table} and \ref{compare-table2}, the the VT constraints allow the decoder to discard a large number of incorrect deletion patterns in Steps and 1 and 3.

As expected, the third step of decoding (determining chunk boundaries) is the most time consuming one in practice. For example, the average wall times for the six decoding steps for setup 6 (for a Matlab implementation on a personal computer) were observed to be: 0.55ms, 0.78ms, 5.5ms, 0.41ms, 0.43ms, 0.10ms.

\subsection{Tradeoffs between redundancy, list sizes, and decoding complexity} \label{subsec:tradeoff}

To get some insight into how the redundancy and the intermediate list sizes decrease with increasing $n$ (for a fixed $k$), consider the choice of parameters $l_1= \alpha_1 k, l_2=\alpha_2 k$ for $\alpha_1, \alpha_2 >0$, and $z = \beta k n_c$ for $\beta \geq 0$.  The bound on the per-symbol redundancy from \eqref{eq:sync_rate3} is
\be
R_{\text{sync}} \leq \frac{\beta}{\alpha_1 \alpha_2 k} +  \frac{ \log(1 + n_c \alpha_1 k)}{\alpha_1 k n_c} + \frac{ \log(1 + n_c \alpha_2 k)}{\alpha_2 k n_c}+\frac{1}{n}.
\label{eq:rsync_s1}
\ee
Let us now consider increasing $n$ and $l_1$ by increasing $\alpha_1$, with $k, n_c, \alpha_2$  fixed. Since $n = n_c l_1l_2 = n_c \alpha_1 \alpha_2 k^2$, $n$ increases linearly with $\alpha_1$.  The first two terms of the per-symbol redundancy in \eqref{eq:rsync_s1} decrease with $\alpha_1$.
Furthermore, the simulation results in Fig. \ref{fig:2figsA} show that the average value of $\abs{\mc{L}_1}$ (list size at the end of Step 1)  decreases as $\alpha_1$ increases.  We therefore expect the average list sizes at the end of Steps 1-4 to decrease with $\alpha_1$  (despite  the upper bound $\binom{k+ \alpha_1 k -1}{k}$ on  $\abs{\mc{L}_1}$ increasing). This in turn allows us to use fewer parity check constraints (smaller value of $\beta$) which helps further  reduce the redundancy as well as decoding complexity.

\begin{table}[t]
\centering
\caption{Complexity, redundancy and error probability changes with $l_1$ when $l_2=4, k=5, n=400$. \vspace{1mm}}\label{tab:tradeoff}
\resizebox{\columnwidth}{!}{
\begin{tabular}{ccccccc} 
\toprule
$(l_1,n_c)$& $R_{\text{sync}}$ &  $\mathbb{P}[\vert\Lc_6\vert >1]$&$\mathbb{E}\vert\Lc_1\vert$& $\mathbb{E}\vert\Lc_3\vert$ &   Decoding Time (s) \\
\midrule   
(4,25) 	& 0.1644 & 0	& 6.8664 &  24.7716  & 0.0378 \\
(5,20) 	& 0.1708 & 0 & 5.8459 &  15.7365  &0.0196 \\
(10,10) & 0.2130 & 0 & 3.0266 & 4.7102   & 0.0081  \\
(20,5)  & 0.2924 & 0 & 1.8694 & 2.1681   & 0.0053 \\
\bottomrule 
\end{tabular}
}
\vspace{-7pt}
\end{table}
\begin{table}[t]
\centering
\caption{Complexity, redundancy and error probability changes with $l_2$ when $l_1=10, k=5, n=400$. \vspace{1mm}}\label{tab:tradeoff2}
\resizebox{\columnwidth}{!}{
\begin{tabular}{ccccccc} 
\toprule
$(l_2,n_c)$& $R_{\text{sync}}$ &  $\mathbb{P}[\vert\Lc_6\vert >1]$&$\mathbb{E}\vert\Lc_1\vert$& $\mathbb{E}\vert\Lc_3\vert$ &   Decoding Time (s) \\
\midrule   
(2,20) 	& 0.1972 & 0	& 2.9710 &  4.6581  & 0.0256 \\
(4,10) 	& 0.2130 & 0 & 2.9706 &  4.5335  &0.0094 \\
(8,5) 	& 0.2536 & 0 & 2.9880 & 2.2901   & 0.0101  \\
(10,4)  & 0.2729 & $1\times 10^{-6}$ & 2.9804 & 2.0346   & 0.0090 \\
\bottomrule 
\end{tabular}
}
\vspace{-7pt}
\end{table}

\definecolor{mycolor1}{rgb}{0.00000,0.44700,0.74100}%
\begin{figure}[t]
\centering
\begin{tikzpicture}[scale=0.65]
\begin{axis}[%
width=4.521in,
height=3.566in,
at={(0.758in,0.481in)},
scale only axis,
xlabel={\large Number of blocks $l_1$},
ylabel={\large Redundancy per symbol},
xmin=2,
xmax=20,
ymin=0.16,
ymax=0.36,
axis background/.style={fill=white},
legend style={legend cell align=left, align=left, draw=white!15!black}
]
\addplot [color=black,line width=2.0pt]
  table[row sep=crcr]{%
2	0.201510516911789\\
3	0.174611143247779\\
4	0.167337373283413\\
5	0.167503383491952\\
6	0.171127526369522\\
7	0.176546320258832\\
8	0.182953765295325\\
9	0.18991875694241\\
10	0.197194058571347\\
11	0.204630512721888\\
12	0.21213473258309\\
13	0.21964667091821\\
14	0.227127022553498\\
15	0.234549810021028\\
16	0.241897847181538\\
17	0.249159871927095\\
18	0.256328683223519\\
19	0.263399901904886\\
20	0.270371129594833\\
};
\addlegendentry{$l_2=2$}
\addplot [color=black,dotted,line width=2.0pt]
  table[row sep=crcr]{%
2	0.167337373283413\\
3	0.161271332952736\\
4	0.164414229655036\\
5	0.170830239863576\\
6	0.178621049407812\\
7	0.187016033773312\\
8	0.195655621666948\\
9	0.204356724425144\\
10	0.21302091494297\\
11	0.221593732729875\\
12	0.230044922288047\\
13	0.238358142674449\\
14	0.24652530749655\\
15	0.254543333059318\\
16	0.262412203553162\\
17	0.270133787122247\\
18	0.277711095150698\\
19	0.285147810908088\\
20	0.292447985966456\\
};
\addlegendentry{$l_2=4$}

\addplot [color=black,mark=x,line width=2.0pt]
  table[row sep=crcr]{%
2	0.167503383491952\\
3	0.165604009827942\\
4	0.170830239863576\\
5	0.178496250072116\\
6	0.187120392949685\\
7	0.196110615410423\\
8	0.205196631875488\\
9	0.214244956855906\\
10	0.22318692515151\\
11	0.231987015665687\\
12	0.240627599163253\\
13	0.249101075959912\\
14	0.257405603419375\\
15	0.265542676601191\\
16	0.273515713761701\\
17	0.281329209095493\\
18	0.288988216470349\\
19	0.296498031642944\\
20	0.303863996174996\\
};
\addlegendentry{$l_2=5$}
\addplot [color=black,mark=triangle,line width=2.0pt]
  table[row sep=crcr]{%
2	0.197194058571347\\
3	0.20362801824067\\
4	0.21302091494297\\
5	0.22318692515151\\
6	0.233477734695746\\
7	0.24365843334696\\
8	0.253637306954882\\
9	0.263380076379745\\
10	0.272877600230904\\
11	0.282132236199627\\
12	0.291151607575981\\
13	0.299945597193152\\
14	0.308524849927341\\
15	0.316900018347252\\
16	0.325081388841096\\
17	0.333078707704299\\
18	0.340901113771966\\
19	0.348557127774969\\
20	0.35605467125439\\
};
\addlegendentry{$l_2=10$}
\end{axis}
\end{tikzpicture}%
\caption{Redundancy per symbol for different values of $l_1$ and $l_2$}
\label{fig:rate}
\end{figure}

We now examine via an example how the choice of the parameters $(n_c, l_1, l_2)$ influence the decoding performance and complexity for fixed $(k,n)$, recalling that $n = n_c l_1 l_2$.
We let $n=400$, $k=5$ and $\beta=0.5$. The code uses $z = \beta k n_c$ binary parity check constraints chosen uniformly at random. Table \ref{tab:tradeoff} shows the effect of increasing $l_1$, with $l_2$ and $n$ fixed. The decoding time decreases with increasing $l_1$ due to fewer valid deletion patterns at the end of the first step. \footnote{In our current implementation, the chunks are all of equal size $n_c$, and therefore $n$ should be divisible by $l_1$ and $l_2$. However, if the binary parity check constraints are drawn uniformly at random (not from a Reed Solomon code), then we do not need equal-sized chunks or blocks. In this case, $n$ does not need to be divisible by $l_1$ and $l_2$.} However, the rate increases with $l_1$ beyond a small value (see Figure \ref{fig:rate}). Therefore there is a tradeoff between rate and complexity when changing $l_1$. Table \ref{tab:tradeoff2} shows the effect of increasing $l_2$ with $(l_1, n)$ fixed. We observe that the effect of $l_2$ on decoding time is not as dominant as $l_1$. This is because of the importance of $l_1$ in reducing number of deletion patterns in the first step. Since the rate is symmetric with respect to $l_1$ and $l_2$, one approach to choose the parameters could be tuning $l_2$ such that it minimizes the rate for the chosen $l_1$. Our experiments shows typically choosing $l_1>k$ and $l_2\approx k$ gives a good tradeoff for values of $n$ and $k$ that we consider in this paper.
\subsection{Comparison with Guess and Check (GC) codes} \label{subsec:GC_comp}

In the GC code, the sequence $X$  of length $n$ (assumed to be a power of 2) is divided into chunks of $\log n$ bits. The encoder's message consists of  $c $ parity-check symbols of a systematic MDS code over $GF(n)$, computed with the information sequence  $X$.   The decoder considers each possible pattern of $k$ deletions,  erases the chunks corresponding to the deletion pattern, and recovers the erased chunks using the MDS decoder. Decoding is successful when the recovered sequence is consistent with each of  the $c$ parity symbols received from the encoder.  The number of  deletion patterns tested by the decoder is ${n/\log n + k -1 \choose k}$, and the MDS decoder run for each deletion pattern has complexity  $O(k^3 n \log n)$ (assuming a Reed-Solomon code). Therefore the decoding complexity of the GC code is 
$O \Big( (n/\log n)^k  \,  k^3 n \log n  \Big)$. In particular, the complexity increases exponentially with the number of deletions $k$. As discussed above, the upper bound on decoding complexity of the multilayer scheme also scales exponentially with $k$. However  the empirical results in Tables \ref{compare-table} and \ref{compare-table2} demonstrate that the `typical' decoding complexity is much lower, due to a large number of deletion patterns being eliminated by the intersecting VT constraints.

\textbf{Example.}  Let us compare the complexity of synchronizing a sequence of length $n=1024$ from $k=8$ deletions with a multilayer code and a GC code. 

\emph{Multilayer code}:  The code parameters were chosen to be $l_1=16$, $l_2=8$, $n_c=8$, and $z=60$ random binary linear constraints. The corresponding per-symbol redundancy  is $R=0.230$. Over  $10^6$ independent simulation trials, all sequences were successfully recovered by the multilayer decoder. From \ref{sec:dec_complexity}, the upper bound for the list size $\abs{\mc{L}_1}$ at the end of Step 1 is
${ k+ l_1 -1 \choose k } = 490,314$.  However, the average list size was observed to be $\mathbb{E} \abs{\mc{L}_1} = 7.27$, which means that on average only $1.4\times 10^{-5}$ of the possible block deletion patterns were forwarded to the subsequent steps. The average list sizes at the end of Steps 3 and 4 were $\mathbb{E}\vert\Lc_3\vert=58.16$ and $\mathbb{E}\vert\Lc_4\vert=2.15$, again much smaller than the upper bounds. The average decoding time per trial (Matlab implementation on a personal computer)  was 0.0614 seconds.

\emph{GC Code}: In the standard construction of the GC code, the sequence is divided into chunks of $\log n  =10$ bits. To synchronize from $k=8$ deletions, ${n/\log n + k -1 \choose k} > 4 \times 10^{11}$. For each of these patterns, the GC decoder has to run an MDS decoder.  For this sequence length $n=1024$, the authors report in \cite{hanna2019guess}   that the GC decoding time is of the order of seconds for $k=3$ deletions, and of the order of minutes for $k=4$ deletions. Due to the prohibitively large number of deletion patterns, decoding is infeasible for $k=8$ with the default choice of chunk length $\log n = 10$.
 
 As suggested in \cite{hanna2019guess}, the number of deletion patterns to be  checked in the GC scheme can be reduced by increasing the chunk length, at the expense of increased redundancy. Consider a chunk length  of $n_c=30$, with $c =(k+1) =9$ parity symbols (which is the minimum required by the decoder). The per-symbol redundancy of the GC code with these parameters is $R= \frac{c n_c}{n} = 0.263$, which is slightly higher than that of the multilayer code above.  The number of deletion patterns to be checked by the GC decoder is  ${ \lceil n/n_c \rceil + k-1 \choose k} = {42 \choose 8} > 10^8$. Since an MDS decoder of complexity  $O(k^3 n \log n)$ has to be run for each of these deletion patterns, GC decoding is still too complex to run on a personal computer.

To summarize, the VT constraints in the multilayer play a crucial role in eliminating a large number of deletion patterns, which makes decoding feasible for a larger number of deletions than the GC scheme.  On the other hand, due to the non-linearity of the VT constraints it is difficult  to obtain sharp analytical bounds on the probability of decoding failure and the typical decoding complexity.

%%%%%%%%%%%%%%%%%%%%%%%%%%%%%%%%%%%%%
%%%%%%%%%%%%%%%%%%%%%%%%%%%%%%%%%%%%%

\section{Guess-based VT decoding}\label{sec:alt-dec}
In this section, we consider an alternative decoder which does not use the parity-check constraints. Recall that the parity-check constraints are used in Step 5 of the decoding algorithm to recover deletions that cannot be directly recovered using the intersecting VT constraints. Here we first characterize such deletions, and then show how they can be often be recovered  using only the VT constraints. Eliminating the parity-check constraints decreases the redundancy, but this comes at the expense of an increased list size.   %{\color{red}(one question is that which one is more efficient. Using random hashes with new decoding or having linear codes and normal decoding?)}

\subsection{Unresolved deletions in step four}
Here we characterize the deletions that cannot be recovered by the iterative algorithm in the step 4 of decoding, for a given chunk-deletion matrix produced in step 3. We use a graph representation for the chunk-deletion matrix to illustrate this. Recall that the chunk deletion matrix $\bm{A}$ consists of entries $\{ a_{ij} \}_{1 \leq i \leq l_1, 1 \leq j \leq l_2}$, where $a_{ij}$ specifies the number of deletions in the $j$th chunk of the $i$th block.

\begin{definition}
Define a bipartite graph $\Gc$ associated with each chunk-deletion matrix $\bm{A}$ with vertex sets $\Bc$ and $\Cc$. Each vertex in $\Bc$ corresponds to a block (row of $\bm{A}$),  and each vertex in $\Cc$ corresponds to a chunk-string (column of $\Ab$). For any non-zero entry $a_{ij}$ of $\bm{A}$, there are $a_{ij}$ edges between the $i$th vertex in $\Bc$ and $j$th vertex in $\Cc$.
\end{definition}
 Figure \ref{fig:graph} shows an example of a chunk-deletion matrix and the corresponding bipartite graph.  Here a vertex $C_j$ represents the $j$th column (chunk-string) of the matrix and $B_i$ represents $i$th row (block). In the following, we will adopt usual definitions of paths and cycles from graph theory. In particular, if there are two edges between two vertices, it is considered a cycle of length $2$.

In step 4, the decoder iteratively corrects deletions by identifying a row or column in $\bm{A}$ with a single one. This corresponds to finding a degree one vertex in the bipartite graph.   When such a vertex is identified, the deletion is corrected and the bipartite graph  updated by removing the edge connected to the degree one vertex. This process is iterated until there are no more degree one edges. In the example in Figure \ref{fig:graph}, $C_5$ is a degree one vertex, indicating that the fifth chunk-string has only one deletion. The deletion corresponding to edge between $B_4$ and $C_5$ is recovered, and this edge is then removed from the graph.   The other deletions remain unresolved as there are no more degree one vertices. The following result determines the graph configurations that result in unrecovered deletions at the end of Step 4.

\begin{figure}[t]
\begin{minipage}{0.45\textwidth}
\begin{equation*}
\bm{A}_0=\begin{bmatrix}
1&1&0&0&0 \\
0&0&0&0&0 \\
1&0&1&0&0 \\
0&1&1&0&1 \\
0&0&0&0&0
\end{bmatrix} \hspace{4mm}
\Longleftrightarrow
\label{matrix-exm}
\end{equation*}
\end{minipage}
\begin{minipage}{0.55\textwidth}
\vspace{10pt}
\begin{tikzpicture}[scale=0.8,shorten >=1pt]
  \tikzstyle{vertex}=[circle,fill=black!25,minimum size=12pt,inner sep=2pt]
  \node[vertex] (G_1) at (0,0) {$B_1$};
  \node[vertex] (G_2) at (0,2)   {$C_1$};
  \node[vertex] (G_3) at (2,0)  {$B_2$};
  \node[vertex] (G_4) at (2,2)   {$C_2$};
  \node[vertex] (G_5) at (4,0)  {$B_3$};
  \node[vertex] (G_6) at (4,2) {$C_3$};
  \node[vertex] (G_7) at (6,0)  {$B_4$};
  \node[vertex] (G_8) at (6,2) {$C_4$};
  \node[vertex] (G_9) at (8,0)  {$B_5$};
  \node[vertex] (G_10) at (8,2) {$C_5$};
  \node[vertex] (G_11) at (10,0)  {$B_6$};
  \node[vertex] (G_12) at (10,2) {$C_6$};
  \foreach \from/\to in {G_1/G_2,G_2/G_5,G_5/G_6,G_6/G_7,G_7/G_4,G_4/G_1,G_7/G_10}
  \draw (\from) -- (\to);
\end{tikzpicture}
\vspace{10pt}
\end{minipage}
\caption{Example of a chunk deletion matrix $\bm{A}_0$ and the corresponding bipartite graph} \label{fig:graph}
\vspace{-8pt}
\end{figure}

\begin{proposition}\label{proposition}
A deletion occurring in the $j$th chunk of the $i$th  block will not be recovered by means of the iterative algorithm if and only if the corresponding edge, $B_iC_j$, in the graph $\Gc$ belongs to a cycle, or belongs to a path between two cycles.
\end{proposition}
\begin{proof}
Consider an unrecovered edge $B_iC_j$ which does not belong to a cycle. As the degree of $B_i$ is greater than one, we can find a vertex other than $C_j$ connected to $B_i$. Similarly, the degree of that vertex is greater than one, hence we can continue this procedure. Since the graph is finite we revisit a vertex which means there is a path from $B_i$ to a cycle. By repeating this argument for $C_j$, we conclude that $B_iC_j$ is in a path which connects two cycles.
\end{proof}
\subsection{Guess-based decoding}
Here we show how to recover the remaining deletions at the end of step 4 of the decoding by guessing bits to break cycles in the bipartite graph. Since there are no parity-check constraints, the per-symbol redundancy is now:
\be R =  \frac{\lceil\log(n_cl_2+1)\rceil}{n_cl_2} + \frac{\lceil \log(n_cl_1+1) \rceil}{n_cl_1},  \label{eq:sync_rate_new} \ee
which is a saving of $z/n$ over the redundancy in \eqref{eq:sync_rate}. 

To motivate the guess-based decoder, consider the matrix $\Ab_0$  and its corresponding graph  in Figure \ref{fig:graph}. After correcting the deletion corresponding to $B_4-C_5$, the remaining deletions form a cycle of length 6 in the graph. If we recover one of the deletions in the cycle, then we can immediately recover all the other deletions  using the iterative algorithm (as there is no other cycle in the graph). We therefore guess the deleted bit (both location and value) in one of the chunks in the cycle. For instance, we can guess the deleted bit in  $C_1$. Since we already know $(n_c-1)$ bits of $C_1$, there are $(n_c+1)$ distinct sequences that can be obtained by inserting one bit into this chunk. The decoder runs the iterative deletion correction algorithm Step 4 for each of these $n_c+1$ obtained sequences.  Since there are no other cycles in the graph, the iterative algorithm will either successfully find all the remaining deletions, or discard the sequence due to the position of the recovered bits  being incompatible with the chunk they are expected to be in (known from $\Ab_0$). The decoder then forwards the remaining sequences, which are now of length $n$, to the sixth step of the decoding algorithm (bypassing the fifth step).

In general, for each unresolved chunk-deletion matrix at the end of Step 4, Proposition \ref{proposition} identifies a minimal set of deletions that need to be guessed in order to resolve all the deletions corresponding to the chunk-deletion matrix. The proposition tells us that it is necessary and sufficient to remove a set of edges such that the remaining graph has no cycles. Hence, the minimum number of edges that need to be removed to make the graph acyclic is equal to the minimum number of bits that need to be guessed. Denote this number by $a^*$. If there are $c$ connected components in the graph with $\alpha_1,\cdots, \alpha_c$ vertices, respectively, then $a^*=e-(\sum_{i=1}^c\alpha_i)+c$, where $e$ is the total number of edges in the graph (total number of deletions).  Since the number of distinct supersequences that can be obtained by inserting $a^*$ bits in a length $(n_c-a^*)$ binary sequence is \cite{levenshtein2001efficient}
\be\label{eq:num:ins}
\sum_{j=0}^{a^*} \binom{n_c}{j}\leq (n_c+1)^{a^*}.
\ee
Using \eqref{eq:num:ins}, $(n_c+1)^{a^*}$ is an upper bound  for the number sequences that need to be guessed. Note that $a^*$ is determined by the specific chunk-deletion matrix (or its bipartite graph).

In our implementation of the algorithm, the decoder chooses one of the edges in a cycle uniformly at random, removes the edge it by guessing a bit in the corresponding chunk, and then performs the iterative algorithm on the updated graph (discarding inconsistent candidates). If any unresolved deletions remain, it chooses another edge from a cycle uniformly at random, and repeats the algorithm until there are no more edges in the graph.

\emph{Numerical simulations}: We present simulation results for the guess-based decoder, for the setups listed in Table \ref{Setup-table}. Setup 8 and 10 are similar to setups 1 and 4 in Section \ref{sec:simulation} respectively, with the only difference being that there is no linear code in setups 8 and 10. The quantity $R'_{\text{sync}}$ in brackets is the higher overall redundancy per symbol when linear codes were used. The performance was recorded over $10^6$ simulation trials.  The first three columns in Table \ref{result-table} show the fraction of trials in which the final list size was exactly 1, 2, and 3 respectively. The fourth column shows the fraction of trials with more than 3 candidates on the final list, and the last column shows the largest list size over all $10^6$ trials.

\begin{table}[t]
\centering
\caption{Number of deletions and code parameters for each setup. \vspace{1mm}}\label{Setup-table}
\begin{tabular}{cccccccc} 
\toprule
  &  $k$ &$n$& $l_1$ & $l_2$ & $n_c$ & $z$  &  $R_{\text{sync}}\  (R'_{\text{sync}})$ \\
\midrule   
Setup 8  &3 &60 & 5 & 3 & 4 & 0\ (4) & 0.583\ (0.650)\\
Setup 9  &3 &60 & 5 & 3 & 6 & 0\ (6) & 0.444\ (0.511) \\
Setup 10 &4 &60 & 5 & 3 & 4 & 0\ (16) & 0.583\ (0.850)  \\
Setup 11 &4 &60 & 5 & 3 & 6 & 0\ (24) & 0.444\ (0.711)\\
\bottomrule 
\end{tabular}
\vspace{-5pt}
\bigskip
\centering
\caption{List size distribution for guess-based decoder.\vspace{1mm}}\label{result-table}
\begin{tabular}{ccccccc} 
\toprule
&$ \vert\Lc_6\vert=1$& $ \vert\Lc_6\vert=2$ & $ \vert\Lc_6\vert=3$  & $ \vert\Lc_6\vert>3$ &  $\max  \vert\Lc_6\vert$ \\
\midrule   
Setup 8       & 92.7\% & 6.3\%  & 0.91\%  & 0.09\%   &6  \\
Setup 9       & 85.9\% & 10.2\% & 2.8\% & 1.1\% &10 \\
Setup 10       & 84.3\% & 12.8\% & 2.3\% & 0.6\%  &13 \\
Setup 11      & 71.9\% & 18.2\% & 6.1\% & 3.8\% &25 \\

\bottomrule 
\end{tabular}
\vspace{-4pt}
\end{table}
Comparing the performance of setups 8 and 10 in Table \ref{result-table} with setups 1 and 4 in Table \ref{Deletion-table} shows that the guess-based iterative decoder allows for smaller rates, but has a much larger probability of having more than one candidate on the final list. Guess-based  decoding is effective if we are willing to tolerate list sizes greater than one with non-negligible probability.

Comparing setups 8 and 9 shows that increasing $n_c$ decreases the redundancy (according to \eqref{eq:sync_rate_new}) but increases the average list size. The reason for this is that when a chunk deletion pattern contains cycles, the number of possible guesses increases with $n_c$. The same effect can be observed by comparing setups 10 and 11. Furthermore, as expected, the list size increases with the number of deletions $k$ as can been seen by  comparing setups 8 and 10 (and also setups 9 and 11).

%%%%%%%%%%%%%%%%%%%%%%%%%%%%%%%%%%%%%%
%%%%%%%%%%%%%%%%%%%%%%%%%%%%%%%%%%%%%%
\section{Synchronizing from a combination of deletions and insertions}\label{sec:ins:dec}

In this section, we use the multilayer code for synchronization when the edits are a combination of insertions and deletions. The code construction and the encoding are unchanged, and as described in Section \ref{sec:encoding}, the message sent by the encoder is of the form $M=[M_1, M_2, M_3]$. We describe the modifications required in the decoding algorithm to recover a combination of up to $k$ deletions and insertions. 
First notice that for the case where we have only insertions we can use nearly the same decoding algorithm used for the deletion only case. (Recall that VT codes can  recover either a single insertion or deletion in a sequence.)

For the case where the edits are a combination of insertions and deletions, assume that the sequence $Y$ is of length $m$  can be obtained from $X$ by $a$ deletions and $b$ insertions where $a+b\leq k$.  (Thus $m=n-a+b$.) We will use a similar six-step decoder for reconstructing $X$.

\subsubsection{Step 1} In this step, we perform a tree search to find the number of insertions and deletions in each  block. The output of this step is a list of block edit patterns  of the form 
\begin{align}
V =\big( (a_1, b_1), \,  (a_2, b_2), \, \cdots, (a_{l_1}, b_{l_1})\big),
\end{align} 
where $a_i$ and $b_i$ are the number of deletions and insertions, respectively, in block $i$ according to the edit pattern. Each valid edit pattern should satisfy
\be \label{eq:ins&del}
\sum_{i=1}^{l_1} (a_i+b_i) \leq k,\quad \sum_{i=1}^{l_1} (a_i-b_i) = n-m.
\ee

The tree search to construct the list of valid block edit patterns proceeds sequentially as follows. Assume that for a given node at level $j$ of the tree corresponds to a total of $d_j$ deletions and $\iota_j$ insertions  in the previous $(j-1)$ blocks. The starting point of the $j$th block is then
\be p_j=(j-1)n_b-d_j+\iota_j+1. \ee
Note that given \eqref{eq:ins&del}, we know that $a_j$ and $b_j$ should satisfy
\be a_j\leq \frac{k+n-m}{2}-d_j, \hspace{4mm} b_j \leq \frac{k+m-n}{2}-\iota_j.\label{eq:ins&del2} \ee
The decoder computes the VT syndrome of the $j$th block, $\syn(Y(p_j:p_j+n_b-1))$. If it does not match with  the correct VT syndrome of block $j$ (which is known from the message sent by the encoder), the possible values for $a_j$ and $b_j$ are all the pairs which satisfy \eqref{eq:ins&del2} and also $a_j+b_j\neq 0$. If $d_j+\iota_j=k$, then we discard the correspond branch of the tree. 

If the VT syndrome of the block matches with the correct VT syndrome, then the possible values for $(a_j, b_j)$ are all the pairs which satisfy \eqref{eq:ins&del2} as well as  $a_j+b_j\neq 1$.

\subsubsection{Step 2} For each valid block edit pattern from step 1, the decoder recovers the edits in the blocks with a single insertion or deletion, i.e. when $a_i+b_i=1$. After recovering the edit in a block, the edit pattern is updated.

\subsubsection{Step 3} The goal of this step is to create a list of chunk-edit matrices, each of dimension $l_1 \times l_2$,  the $(i,j)$ entry of the matrix is a pair $(a_{ij}, b_{ij})$. Here $a_{ij}, b_{ij}$ denote the number of deletions and insertions, respectively, in the $j$th chunk of $i$th block. Similar to step 3 of decoding in the deletion-only case, we construct these chunk-edit matrices via a tree search for each block edit pattern of the form $\left( (a_1, b_1), \,  (a_2, b_2), \, \cdots, (a_{l_1}, b_{l_1})\right)$. This is done sequentially as follows. 

For each node at level $j$ of the tree, the decoder knows $a_{ih}$ and $b_{ih}$ for all $i$ and $h<j$. 
Thus it knows the starting position of the $j$th chunk of each block, and can therefore form the $j$th chunk-string and compute its VT syndrome. This computed VT syndrome  is compared with the correct syndrome of $j$th chunk-string (known from the encoder's message). There are two possibilities:

\begin{enumerate}
\item If the VT syndrome of the $j$th chunk-string matches  the correct syndrome, the possible values for $a_{ij}$ and $b_{ij}$ are all non-negative integers that satisfy $\sum_{i=1}^{l_1} (a_{ij}+b_{ij})\neq 1$ and also:
\be 
a_{ij}\leq a_i-\sum_{h=1}^{j-1} a_{ih} \hspace{5mm} \text{and} \hspace{5mm} b_{ij}\leq b_i-\sum_{h=1}^{j-1} b_{ih}. 
\label{chunk-tree-res}
\ee
\item If the VT syndrome of the $j$th chunk-string does not match  the correct syndrome, the possible values for $a_{ij}$ and $b_{ij}$ are all non-negative integers that satisfy \eqref{chunk-tree-res}, and also $\sum_{i=1}^{l_1} (a_{ij}+b_{ij})\neq 0$. The node will be discarded if
\be a_i=\sum_{h=1}^{j-1} a_{ih} \hspace{3mm} \text{and} \hspace{3mm}  b_i=\sum_{h=1}^{j-1} b_{ih}, \hspace{3mm} \text{for} \hspace{3mm} 1\leq i\leq l_1.
\ee
\end{enumerate}

\begin{table}[t]
\centering
\caption{List size after each step when there are both insertions and deletions. \vspace{1mm}}\label{InDel-table}
\resizebox{\columnwidth}{!}{
\begin{tabular}{cccccccc} 
\toprule
&$\mathbb{E}\vert\Lc_1\vert$& $\mathbb{E}\vert\Lc_3\vert$ & $\mathbb{E}\vert\Lc_4\vert$  & $\mathbb{E}\vert\Lc_6\vert$ &  $\max \vert\Lc_6\vert$&  $\mathbb{P}[\vert\Lc_6\vert >1]$ \\
\midrule   
Setup 1 	  & 2.96 & 3.44	& 2.12 & 1.004 & 7 &$4.215\times 10^{-4}$  \\
Setup 2 	  & 2.96 & 3.44 & 2.12 & 1.000 &2 &$1.3\times 10^{-5}$  \\
Setup 3 	  & 2.96 & 3.44 & 2.12 & 1.000  &2 &$5\times 10^{-6}$  \\
Setup 4      & 7.78 & 17.66 & 5.95 & 1.000  &2 &$4\times 10^{-6}$  \\
Setup 5      & 86.29 & 782.38& 22.5& 1      & 1&0   \\
Setup 6      & 82.73 & 254.06& 15.08& 1      & 1&0  \\
Setup 7      & 210.74 & 1523.0& 34.41& 1      & 1&0  \\
\bottomrule 
\end{tabular}
}
\label{tab:indel_sims}
\end{table}

\subsubsection{Steps 4 to 6} The last three steps are very similar to the deletion only case. In step 4, we iteratively use the VT decoder to recover any single deletion in blocks or chunk-strings. Similarly to the deletion only case, we will discard a candidate if the recovered bit lies in a wrong chunk. In step 5, we replace any chunk which still contains edits with $n_c$ erasures, and use the parity-check constraints to recover erasures; any inconsistent candidate will be discarded. Finally, at the sixth step we check all constraints for the remaining candidates and output all compatible sequences.

\emph{Numerical simulations}: We evaluated the performance of the decoder for the seven setups in Table \ref{Setup-table}.  For each $k$,  the number of deletions was chosen to be an integer $d$ between $0$ and $k$ uniformly at random. The number of insertions was then  $(k-d)$. 
 Table \ref{tab:indel_sims} shows the average list size for each each of the setups, in the different steps of the decoding. As expected, with a combination of insertions and deletions, the number of valid block edit patterns (in step 1) and chunk edit matrices (in Step 3) are larger than the deletion-only case. 
 As the decoding complexity of each step depends on the list size at the end of the previous step, the average decoding complexity is also higher than the deletion-only case.
  However, we observe that the increase in the final list size compared to the deletion-only case (Table \ref{Deletion-table}) is negligible. This is because a large number of the edit patterns are inconsistent with the intersecting VT constraints and the parity-check constraints.

\begin{figure}[tp]
\centering
\begin{tikzpicture}[scale=1.2]
    \tikzstyle{every node}=[draw,shape=circle];
    \path (1,1) node (v1) {$v_1$};
    \path (2,1) node (v2) {$v_2$};
    \path (3,1) node (v3) {$v_3$};
    \path (4,1) node (v4) {$v_4$};
    \path (5,1) node (v5) {$v_5$};
    \path (6,1) node (v6) {$v_6$};
    \path (7,1) node (v7) {$v_7$};
    \path (8,1) node (v8) {$v_8$};
    \tikzstyle{every node}=[draw,shape=rectangle];
    \path (1.3,3) node (b1) {$B_1$};
    \path (2.6,3) node (c1) {$C_1$};
    \path (3.9,3) node (s1) {$T_1$};
    \path (5.2,3) node (s2) {$T_2$};
    \path (6.5,3) node (c2) {$C_2$};
    \path (7.8,3) node (b2) {$B_2$};
        \draw (v1) -- (b1)
              (v1) -- (c1)
              (v1) -- (s1)
              (v2) -- (b1)
              (v2) -- (c1)
              (v2) -- (s2)
              (v3) -- (b1)
              (v3) -- (c2)
              (v3) -- (s1)
              (v4) -- (b1)
              (v4) -- (c2)
              (v4) -- (s2)
              (v5) -- (b2)
              (v5) -- (c1)
              (v5) -- (s1)
              (v6) -- (b2)
              (v6) -- (c1)
              (v6) -- (s2)
              (v7) -- (b2)
              (v7) -- (c2)
              (v7) -- (s1)
              (v8) -- (b2)
              (v8) -- (c2)
              (v8) -- (s2);
\end{tikzpicture}
\caption{Factor graph representation of a three-layer code.} \label{fig:factorg}
\vspace{-5pt}
\end{figure}

%%%%%%%%%%%%%%%%%%%%%%%%%%%%%%%%%

\section{Discussion and future work} \label{sec:conc}

In this work we introduced a new method for one-way synchronization of binary sequences based on a combination of  intersecting VT constraints and linear parity-check constraints. We showed that the intersecting VT constraints enable a iterative decoding procedure which alternates between identifying compatible edit patterns,  and correcting subsequences  indicated by these patterns as having a single edit. 

\emph{Generalizing the two-layer construction}:
%\label{sec:gen_mult_layer}
The code construction based on two layers of intersecting VT constraints can be generalized in many ways.  First, it can be extended to sequences over non-binary alphabets with size $q >2$ by using the $q$-ary VT codes proposed by Tenengolts \cite{Tenengolts84}. 
The construction can also be generalized to include multiple layers of intersecting VT constraints. We illustrate the idea with an example of a three-layer construction. Consider a sequence  $X=[v_1,v_2,\cdots,v_8]$ consisting of eight chunks of length $n_c$ each. The message consists of syndromes corresponding to three kinds of intersecting VT constraints, defined as follows. The two block constraints, $B_1$ and $B_2$, are the VT syndromes of $[v_1,v_2,v_3,v_4]$ and $[v_5,v_6,v_7,v_8]$.  The two chunk-string constraints $C_1$ and $C_2$, are the VT syndromes of $[v_1,v_2,v_5,v_6]$ and $[v_3,v_4,v_7,v_8]$. The third set of constraints, $T_1$ and $T_2$,  are the VT syndromes of $[v_1,v_3,v_5,v_7]$ and $[v_2,v_4,v_6,v_8]$. Figure \ref{fig:factorg} illustrates the three sets of constraints using a factor graph, with the circles and squares  representing the chunks and constraints, respectively.  The decoding algorithm for a such a construction is a straightforward extension of that in Section \ref{sec:decoding}: we identify the compatible chunk edit patterns via a tree search, and then using the VT constraints to iteratively solve for sub-sequences with a single deletion. 

Extending this idea further, we could consider an $L$-layer construction with $L = \Theta(\log n)$ layers, $l_1= l_2 = \ldots = l_L=2$, $n_c= \log n$, and $z=kn_c$ binary parity check constraints. Such a construction would have an overall redundancy of $k \log n + 2L \log(\frac{n}{2}+1)$, which is near-optimal. Moreover, since each layer has only two VT constraints  the number of sequences compatible with each layer is at most $k$. Developing an iterative decoding algorithm to recover the chunks from these constraints, and investigating the trade-offs between redundancy, list-size, and decoding complexity  is an interesting direction for future research.
 
Another direction for future work is to use the multilayer code construction for  communication over the deletion channel. Such a  channel code will consist of all sequences with a specified set of values for the intersecting VT and parity-check constraints. The decoding algorithm  is essentially the same as that described in Section \ref{sec:decoding}, however constructing an efficient encoder  for this channel code is an open question.

%%%%%%%%%%%%%%%%%%%%%%%%%%%%%%%%%%%
%%%%%%%%%%%%%%%%%%%%%%%%%%%%%%%%%%%
\vspace{1in}
\appendix
\subsection{Proof of Lemma \ref{eq-lemma}} \label{App-lemma1}
\begin{proof}
Define the function $p(x)=\binom{x+l_2-1}{l_2-1}$. We note that $p$ is a polynomial of degree $(l_2-1)$, and $p(a_i)$ is the number of non-negative integer solutions to the equation $\sum_{j=1}^{l_2} a_{ij} = a_i$.

We first show that $p(x)p(y)\leq p(\frac{x+y}{2})^2$ for any two  positive real numbers $x,y$. To show this, we need to prove
\be 
\prod_{i=1}^{l_2-1} (y+i)(x+i)\leq \prod_{i=1}^{l_2-1}\Big(\frac{x+y}{2}+i \Big)^2,
\ee
which clearly follows from $xy\leq (\frac{x+y}{2})^2$. Now if we define $g(x)\triangleq \ln (p(x))$, we have $g(x)+g(y)\leq 2g(\frac{x+y}{2})$ which means $g$ is mid-point concave, and since it is continuous, it is generally concave. Hence, we have $g(x_1)+g(x_2)+\cdots+g(x_n)\leq ng(\sum_{i=1}^n x_i/n)$ for any integer $n$ and positive $x_i$'s. Therefore, we have
\be 
p(x_1)p(x_2)\cdots p(x_n)\leq p\left(\frac{\sum_{i=1}^n x_i}{n}\right)^n. 
\ee
Choosing $n=s$ and $x_i=a_i$ yields the result.
\end{proof}

\subsection{Derivative of \eqref{step3-upper}} \label{app:der}
Here we show that the derivative of $f(s)=\binom{k/s+l_2-1}{l_2-1}^s$ with respect to $s$ is positive for $s>0$. We write $f(s)=h(s; l_2)^s$, where
\be h(s; l_2) = \binom{k/s+l_2-1}{l_2-1}. \ee 
Hence, we have:
\begin{align}
\frac{d\ln f(s)}{ds}&=\frac{1}{f(s)}f'(s) =\ln (h(s; l_2)) \, + \, \frac{s}{h(s; l_2)} h'(s; l_2).
\end{align} 
Therefore, to prove $df/ds>0$ for $s>0$ we need to show that 
\be \ln (h(s; l_2))+\frac{s}{h(s; l_2)}h'(s; l_2)>0.
 \label{eq:der:pos}
 \ee
From  the definition in \eqref{def-binom}, we can write $h(s; l_2) = p(s^{-1}; l_2)/(l_2-1)!$, where 
\be p(s; l_2)=(ks+1)(ks+2)\cdots(ks+l_2-1). \label{eq:psl2_def} \ee
Using this  in \eqref{eq:der:pos}, we need to show that 
\begin{align}
\ln (h(s; l_2))&>\frac{p'(s^{-1}; l_2)}{sp(s^{-1}; l_2)}.
\label{eq:der:pos1}
\end{align} 
We prove \eqref{eq:der:pos1} by induction on $l_2$. For $l_2=2$, we need to show that
\be \ln(ks^{-1}+1)>\frac{k}{(k+s)}\label{eq:ind:base}\ee
Letting $x = ks^{-1}$, then we can rewrite \eqref{eq:ind:base} as $\ln(x+1)>\frac{x}{(x+1)}$,
which holds for all $x>0$.
Assuming that \eqref{eq:der:pos1} holds for $l_2$, we prove it for $l_2+1$. We have 
\begin{align}
\ln \left(h(s; l_2+1)\right)&=\left(\ln(ks^{-1}+l_2)-\ln(l_2)\right)+ \ln \left(h(s; l_2)\right) \\
&>\left(\ln(ks^{-1}+l_2)-\ln\left(l_2\right)\right)+\frac{p'(s^{-1}; l_2)}{sp(s^{-1}; l_2)},\label{eq:ind:left}
\end{align}
where the inequality holds by the induction hypothesis. 

Using \eqref{eq:psl2_def}, we have 
\be
\frac{p'(s; l_2+1)}{p(s; l_2+1)} = \frac{d\ln p(s; l_2+1)}{ds} = \sum_{i=1}^{l_2} \frac{k}{i+ ks}.
\ee
Therefore, 
\be 
\frac{p'(s^{-1}; l_2+1)}{sp(s^{-1}; l_2 +1)} =\frac{k}{k+ l_2 s} + \frac{p'(s^{-1}; l_2)}{sp(s^{-1}; l_2 )}. 
\label{eq:ind:fin}
\ee
Comparing the RHS  of \eqref{eq:ind:left} and \eqref{eq:ind:fin} we have to prove that  
\be \ln\left(\frac{ks^{-1} + l_2 }{ l_2}\right)>\frac{k}{k+ l_2 s}.\ee
Letting $x = ks^{-1}/l_2$, this is equivalent to showing that  $\ln\left(x +1 \right)>\frac{x}{x+ 1}$.
This holds for all $x>0$, which completes the proof.

\vspace{30 mm}

\end{document}